\documentclass[12pt]{article}
 \usepackage[cp1251]{inputenc} 
 \usepackage[english, russian]{babel} 
 \usepackage{amsmath, latexsym, amssymb, bm, array, graphics, eucal,
 amsfonts,}
 \makeatletter

\newcommand{\const}{\mathop{\rm const\, }}

\renewcommand{\div}{\mathop{\rm div}}
 \makeatother
 \renewcommand{\baselinestretch}{1.2}
\usepackage[dvips]{graphicx}

\begin{document}
\thispagestyle{empty}
\large
\renewcommand{\abstractname}{Abstract}
\renewcommand{\refname}{\begin{center}
 REFERENCES\end{center}}
\newcommand{\mc}[1]{\mathcal{#1}}
\newcommand{\E}{\mc{E}}
\makeatother

\begin{center}
\bf  Nonlinear interaction of electromagnetic field with quantum plasma
\end{center} \medskip
\begin{center}
  \bf A. V. Latyshev\footnote{$avlatyshev@mail.ru$} and
  A. A. Yushkanov\footnote{$yushkanov@inbox.ru$}
\end{center}\medskip

\begin{center}
{\it Faculty of Physics and Mathematics,\\ Moscow State Regional
University, 105005,\\ Moscow, Radio str., 10A}
\end{center}\medskip

\begin{abstract}
The analysis of nonlinear interaction of transversal
electromagnetic field with quantum collisionless plasma is carried out.
Formulas for calculation electric current in quantum collisionless plasma
at any temperature are deduced.
It has appeared, that the nonlinearity account leads to occurrence
of the longitudinal electric current directed along a wave vector.
This second current is orthogonal to the known transversal classical current,
received at the classical linear analysis.
The case of degenerate electronic plasma is considered.
It is shown, that for degenerate plasmas the electric current is
calculated under the formula, not containing quadratures.
In this formula we have allocated known Kohn's singularities (W. Kohn, 1959).


\medskip

electron gas
\end{abstract}

\begin{center}
\bf  Introduction
\end{center}
Dielectric permeability in quantum plasma was studied by many
aut\-hors \cite{Klim} -- \cite{Dressel}.
Dielectric permeability is one of the major plasma charac\-te\-ris\-tics.

This quantity is necessary for the description of skin-effect
\cite{Gelder}, for the analysis surface plasmons \cite{Fuchs},
for descriptions of process of propagation and attenuation of the
transversal plasma oscillations \cite{Shukla2}, for studying of
the mechanism of penetration electromagnetic waves in plasma
\cite{Shukla1}, and for the analysis of other problems in the
plasma physics \cite{Brod} -- \cite {Lat4}.

Let us notice, that for the first time in work \cite{Klim} the formula
for cal\-cu\-la\-tion of longitudinal dielectric permeability into
quantum plasma has been deduced. Then the same formula has been deduced
and in work \cite{Lin}.

In the present work formulas for calculation electric current
into quantum collisionless plasma  at any temperature (at any
degrees of degeneration of the electronic gas)  are deduced.

Here the approach developed by Klimontovich Silin \cite{Klim} is
generalized.

At the solution of Schr\"odinger equation we consider and in expansion
of distribution Wigner  function, and in Wigner---Vlasov integral
expan\-sion the quantities proportional
to square of potential of an external electro\-mag\-ne\-tic field.

It has appeared, that electric current expression consists of two sum\-mands.
The first summand, linear on vector potential, is
known classical expression of an electric current.
This electric current is directed along vector potential electromagnetic
field. The second summand represents itself an electric current,
which is proportional to the square
vector potential of electromagnetic fields. The second current
it is perpendicular to the first and it is directed along the  wave
vector.
Occurrence of the second current comes to light the spent account
nonlinear character
interactions of an electromagnetic field with quantum plasma.

For the case of degenerate quantum plasma expression of the electric
current, not containing quadratures, is received.
At the deducing of this expression
Landau' rule for calculation singular integrals is used.
At use of this rule calculation
these integrals containing a pole on the real axis,
it is carried out by means of integration on infinitesimal
semi-circles in the bottom semi-plane with the centre in this pole.

\begin{center}
  {\bf 1. Kinetic equation for Wigner function}
\end{center}

Let uss consider Shr\"odinger equation which has been written down
for a particle in an electromagnetic field on a density matrix
$\rho$:

$$
i\hbar \dfrac{\partial \rho}{\partial t}=H\rho-{H^*}'\rho.
\eqno{(1.1)}
$$

Here $H$ is the Hamilton operator, $H^*$ is the compex conjugate operator to
$H$, ${H^*}'$ is the compex conjugate operator to
$H$, which operates on the shaded spatial variables $\mathbf{r}'$.

The operator of Hamilton of the free particle which is in the field
of the scalar potential $U $ and in the field of vector potential
$\mathbf{A}$, looks like:
$$
H=\dfrac{(\mathbf{p}-({e}/{c})\mathbf{A})^2}{2m}+eU=
$$
$$=
\dfrac{\mathbf{p}^2}{2m}-
\dfrac{e}{2mc}(\mathbf{p}\mathbf{A}+\mathbf{A}\mathbf{p})
+\dfrac{e^2}{2mc^2}\mathbf{A}^2+eU.
\eqno{(1.2)}
$$\smallskip

Here $\mathbf{p}$ is the momentum operator, $\mathbf{p}=-i\hbar \nabla$,
$e$ is the electron charge, $m$ is the electron mass, $c$ is the
speed of light.

We will rewrite the operator of Hamilton (1.2) in the explicit form
$$
H=-\dfrac{\hbar^2}{2m}\triangle+\dfrac{ie\hbar}{2mc}\Big(2\mathbf{A}\nabla
+\nabla\mathbf{A}\Big)+
\dfrac{e^2}{2mc^2}\mathbf{A}^2+eU.
\eqno{(1.3)}
$$

The complex conjugate  to  the $H$  operator $H^*$ according to
(1.3) looks like
$$
H^*=-\dfrac{\hbar^2}{2m}\triangle-\dfrac{ie\hbar}{2mc}
\Big(2\mathbf{A}\nabla+\nabla\mathbf{A}\Big)+
\dfrac{e^2}{2mc^2}\mathbf{A}^2+eU.
$$ \smallskip

Operators $H$ and ${H^*}'$,  calculated on the density matrix,
look like
$$
H\rho=-\dfrac{\hbar^2}{2m}\Delta \rho+\dfrac{ie\hbar}{2mc}
\Big(2\mathbf{A}\nabla \rho+\rho\nabla\mathbf{A}\Big)+
\dfrac{e^2}{2mc^2}\mathbf{A}^2\rho+eU \rho
\eqno{(1.4)}
$$
and
$$
{H^*}'\rho=-\dfrac{\hbar^2}{2m}\Delta'\rho-\dfrac{ie\hbar}{2mc}
\Big(2\mathbf{A'}\nabla'\rho+\rho\nabla'\mathbf{A}\Big)+
\dfrac{e^2}{2mc^2}\mathbf{A'}^2\rho+eU' \rho.
\eqno{(1.5)}
$$

Operators $\nabla$ and $\Delta$ in (1.4) and (1.5) take action on
not shaded spatial variable matrixes of density, i.e.
$ \nabla =\nabla_{\mathbf {R}}$, $ \Delta =\Delta_{\mathbf {R}} $.
In the operator ${H^*}'$ it is necessary to replace operators
$ \nabla =\nabla_{\mathbf {R}} $ and $ \Delta =\Delta_{\mathbf {R}} $
on operators
$ \nabla'\equiv\nabla_{\mathbf {R}'}$ and
$ \Delta'\equiv\Delta_{\mathbf {R}'}$ accordingly.

Besides,  here the following designations are entered

$$
\mathbf{A'}\equiv \mathbf{A}(\mathbf{R'},t),
\quad
\quad U'\equiv U(\mathbf{R'},t).
$$

Let's find the right part of the equation (1.1), i.e. the difference
equalities (1.4) and (1.5) $: H \rho - {H^*}'\rho $. According
to equalities (1.4) and (1.5) it is had
$$
H\rho-{H^*}'\rho=-\dfrac{\hbar}{2m}\Big(\Delta \rho-\Delta'\rho\Big)+
\hspace{4cm}
$$
$$
\hspace{1cm}+\dfrac{i e \hbar}{2mc}
\Big[2\Big(\mathbf{A}\nabla \rho+
\mathbf{A'}\nabla'\rho\Big)+\rho\Big(\nabla \mathbf{A}+
\nabla'\mathbf{A}\Big)\Big]+
$$
$$
\hspace{2cm}
+\dfrac{e^2}{2mc^2}\Big[\mathbf{A}^2(\mathbf{R},t)-\mathbf{A}^2(
\mathbf{R'},t)\Big]+
e[U(\mathbf{R},t)-U(\mathbf{R'},t)]\rho.
$$\medskip

Connection  between the density matrix
$\rho(\mathbf{r},\mathbf{r}',t)$ and
Wigner' function \cite{Hillery} $f(\mathbf{r},\mathbf{p},t)$
is given by inverse and direct Fourier transformation
$$
f(\mathbf{r},\mathbf{p},t)=\int
\rho(\mathbf{r}+\dfrac{\mathbf{a}}{2},\mathbf{r}-
\dfrac{\mathbf{a}}{2},t)e^{-i\mathbf{p}\mathbf{a}/\hbar}d^3a,
$$
$$
\rho(\mathbf{R},\mathbf{R}',t)=\dfrac{1}{(2\pi \hbar)^3}
\int f(\dfrac{\mathbf{R}+\mathbf{R}'}{2}, \mathbf{p},t)
e^{i\mathbf{p}(\mathbf{R}-\mathbf{R}')/\hbar}d^3p.
$$

Wigner' function is analogue of  distribution function for quantum
systems. It is widely used in the diversified questions
physicists.

Substituting representation of the density matrix  through Wigner'
function in Shr\"odinger equation on the matrix of density (1.1), we receive

$$
i\hbar\dfrac{\partial \rho}{\partial t}=H\left\{\dfrac{1}{(2\pi\hbar)^3}
\int f(\dfrac{\mathbf{R}+\mathbf{R}'}{2},\mathbf{p'},t)
e^{i\mathbf{p'}(\mathbf{R}-\mathbf{R}')/\hbar}\,d^3p'\right\}-
$$
$$
\hspace{1.8cm}-{H^*}'\left\{\dfrac{1}{(2\pi\hbar)^3}
\int f(\dfrac{\mathbf{R}+\mathbf{R}'}{2},
\mathbf{p'},t)
e^{i\mathbf{p'}(\mathbf{R}-\mathbf{R}')/\hbar}\,d^3p'\right\}.
$$

We will take advantage of the equalities written above. Thus the right
part the previous equation we will present in an explicit form and
as a result it is received following equation

$$
i\hbar \dfrac{\partial \rho}{\partial t}=
\dfrac{1}{(2\pi \hbar)^3}\int\Big\{-\dfrac{i\hbar}{m}
\mathbf{p'}\nabla f +
\dfrac{ie\hbar}{2mc} \Big[\div{\mathbf{A}(\mathbf{R},t)}+
\div{\mathbf{A}(\mathbf{R}',t)}\Big]f+
$$
$$
+\dfrac{ie\hbar}{2mc}\Big[\mathbf{A}(\mathbf{R},t)+
\mathbf{A}(\mathbf{R}',t)\Big]
\nabla f-\dfrac{e}{mc}\Big[\mathbf{A}(\mathbf{R},t)-
\mathbf{A}(\mathbf{R}',t)\Big]
\mathbf{p'}f+
$$
$$
+\dfrac{e^2}{2mc^2}\big[\mathbf{A}^2(\mathbf{R},t)-
\mathbf{A}^2(\mathbf{R}',t)\big]f+$$$$+
e\big[U(\mathbf{R},t)-U(\mathbf{R}',t)\big]f\Big\}
e^{i\mathbf{p'}(\mathbf{R}-\mathbf{R}')/\hbar}
d^3p'.
\eqno{(1.6)}
$$

We put in equation (1.6)
$$
\mathbf{R}=\mathbf{r}+\dfrac{\mathbf{a}}{2},\qquad
\mathbf{R}'=\mathbf{r}-\dfrac{\mathbf{a}}{2}.
$$

Then in this equation
$$
 f\Big(\dfrac{\mathbf{R}+\mathbf{R}'}{2}, \mathbf{p'},t\Big)
e^{i\mathbf{p'}(\mathbf{R}-\mathbf{R}')/\hbar}=f(\mathbf{r}, \mathbf{p'},t)
e^{i\mathbf{p'\,a}/\hbar}.
$$

We multiply the equation (1.6) on $e^{-i\mathbf{p}\mathbf {a}/\hbar} $ and
we integrate on $ \mathbf{a} $. Then we divide both parts of the equation on
$i\hbar $.  As a result we receive
$$
\dfrac{\partial f}{\partial t}=
\iint\Bigg\{-\dfrac{\mathbf{p'}}{m}\nabla f +
\dfrac{e}{2mc}
\Big[\mathbf{A}(\mathbf{r}+\dfrac{\mathbf{a}}{2},t)+
\mathbf{A}(\mathbf{r}-\dfrac{\mathbf{a}}{2},t)\Big]\nabla
f+ \medskip
$$
$$
+\dfrac{ie}{mc\hbar}\Big[
\mathbf{A}(\mathbf{r}+\dfrac{\mathbf{a}}{2},t)
-\mathbf{A}(\mathbf{r}-\dfrac{\mathbf{a}}{2},t)\Big]\mathbf{p'} f+
$$\medskip
$$+
\dfrac{e}{2mc}\Big[\div{\mathbf{A}(\mathbf{r}}+\dfrac{\mathbf{a}}{2},t)+
\div{\mathbf{A}(\mathbf{r}-\dfrac{\mathbf{a}}{2},t)}\Big]f-
$$\medskip
$$
-\dfrac{ie^2}{2mc^2\hbar}\Big[\mathbf{A}^2(\mathbf{r}+
\dfrac{\mathbf{a}}{2},t)-
\mathbf{A}^2(\mathbf{r}-\dfrac{\mathbf{a}}{2},t)\Big]f-
$$\medskip
$$
-\dfrac{ie}{\hbar}\Big[U(\mathbf{r}+\dfrac{\mathbf{a}}{2},t)-
U(\mathbf{r}-\dfrac{\mathbf{a}}{2},t)\Big]f\Bigg\}
e^{i(\mathbf{p'}-\mathbf{p})\mathbf{a}/\hbar}
\dfrac{d^3a\,d^3p'}{(2\pi\hbar)^3}.
\eqno{(1.7)}
$$

In the left part of the equation (1.7) is
$f=f (\mathbf{r},\mathbf {p},t)$, under integral is
$f=f(\mathbf{r},\mathbf{p'},t)$.

We consider the integral
$$
\iint\mathbf{p'}(\nabla f)e^{i(\mathbf{p'}-\mathbf{p})\mathbf{a}/\hbar}
\dfrac{d^3a\,d^3p'}{(2\pi\hbar)^3}=
\nabla\iint\mathbf{p'} f e^{i(\mathbf{p'}-\mathbf{p})\mathbf{a}/\hbar}
\dfrac{d^3a\,d^3p'}{(2\pi\hbar)^3}=
$$
$$
=\nabla\int\mathbf{p'}f
\delta(\mathbf{p'}-\mathbf{p})d\,\mathbf{p'}=\mathbf{p}\nabla
f(\mathbf{r},\mathbf{p},t).
$$

Two other equalities are similarly checked
$$
\iint\dfrac{e}{mc}\mathbf{A}(\mathbf{r},t)(\nabla f(\mathbf{r},
\mathbf{p}',t))
e^{i(\mathbf{p'}-\mathbf{p})\mathbf{a}/\hbar}
\dfrac{d^3a\,d^3p'}{(2\pi\hbar)^3}=$$$$=
\dfrac{e}{mc}\mathbf{A}(\mathbf{r},t) \nabla f(\mathbf{r},\mathbf{p},t),
$$
and
$$
\iint\dfrac{e}{mc}(\div{\mathbf{A}(\mathbf{r},t)}) f(\mathbf{r},
\mathbf{p}',t)
e^{i(\mathbf{p'}-\mathbf{p})\mathbf{a}/\hbar}
\dfrac{d^3a\,d^3p'}{(2\pi\hbar)^3}=$$$$=
\dfrac{e}{mc}(\div{\mathbf{A}}(\mathbf{r},t)) f(\mathbf{r},\mathbf{p},t).
$$

Then the equation (1.6) we can rewrite in the form

$$
\dfrac{\partial f}{\partial t}+\dfrac{1}{m}
\Big(\mathbf{p}-\dfrac{e}{c}\mathbf{A}\Big)\nabla f-\dfrac{e}{mc}
(\div{\mathbf{A}(\mathbf{r},t)})f(\mathbf{r},\mathbf{p},t)=
W[f].
\eqno{(1.8)}
$$

In equation (1.8) symbol $W[f]$ is the Wigner---Vlasov' integral
(func\-tio\-nal), defined by equality
$$
W[f]=
\iint\left\{
\dfrac{e}{2mc}
\Big[\mathbf{A}(\mathbf{r}+\dfrac{\mathbf{a}}{2},t)+
\mathbf{A}(\mathbf{r}-\dfrac{\mathbf{a}}{2},t)-
2\mathbf{A}(\mathbf{r},t)\Big]\nabla f\right.+
$$\smallskip
$$
+\dfrac{ie}{ mc\hbar}\Big[
\mathbf{A}(\mathbf{r}+\dfrac{\mathbf{a}}{2},t)
-\mathbf{A}(\mathbf{r}-\dfrac{\mathbf{a}}{2},t)\Big]\mathbf{p'}
f+
$$\smallskip
$$
+\dfrac{e}{2mc}\Big[\div{\mathbf{A}(\mathbf{r}+
\dfrac{\mathbf{a}}{2},t)
}+\div{\mathbf{A}(\mathbf{r}-\dfrac{\mathbf{a}}{2},t)}-
2\div{\mathbf{A}(\mathbf{r},t)}\Big]f-
$$\smallskip
$$-
\dfrac{i e^2}{2 mc^2\hbar}\Big[\mathbf{A}^2(\mathbf{r}+
\dfrac{\mathbf{a}}{2},t)-
\mathbf{A}^2(\mathbf{r}-\dfrac{\mathbf{a}}{2},t)\Big]f-
$$\smallskip
$$
-\left. \dfrac{ie}{\hbar}\Big[U(\mathbf{r}+\dfrac{\mathbf{a}}{2},t)-
U(\mathbf{r}-\dfrac{\mathbf{a}}{2},t)\Big]f\right\}
e^{i(\mathbf{p'}-\mathbf{p})\mathbf{a}/\hbar}
\dfrac{d^3a\,d^3p'}{(2\pi\hbar)^3}.
\eqno{(1.9)}
$$

Energy of a particle is equal
$$
\E=\E(\mathbf{r}, \mathbf{p},t)=\dfrac{1}{2m}\Big(\mathbf{p}-
\dfrac{e}{c}\mathbf{A}(\mathbf{r},t)\Big)^2+eU(\mathbf{r},t).
$$

Therefore velocity of a particle $ \mathbf{v} $ is equal
$$
\mathbf{v}=\mathbf{v}(\mathbf{r}, \mathbf{p},t)=
\dfrac{\partial \E}{\partial \mathbf{p}}=
\dfrac{1}{m}\Big(\mathbf{p}-\dfrac{e}{c}\mathbf{A}\Big),
$$
besides,
$$
\nabla \mathbf{v}=-\dfrac{e}{mc}\div{\mathbf{A}}.
$$

Hence, the left part of the equation (1.9) is equal
$$
\dfrac{\partial f}{\partial t}+\dfrac{1}{m}
\Big(\mathbf{p}-\dfrac{e}{c}\mathbf{A}\Big)\nabla f-f\dfrac{e}{mc}
\div{\mathbf{A}}=
\dfrac{\partial f}{\partial t}+\mathbf{v}\nabla f+f\nabla \mathbf{A}=
$$
$$
=\dfrac{\partial f}{\partial t}+\nabla(\mathbf{v}f).
$$

Therefore the equation (1.9) can be rewritten in the form, which is
standard form for the pransport theory

$$
\dfrac{\partial f}{\partial t}+\nabla (\mathbf{v}f)=W[f].
\eqno{(1.10)}
$$

Let $f^{(0)}$ is the locally equilibrium Fermi---Dirac' distribution,
$$
f^{(0)}=\Big[1+\exp\Big(\dfrac{\E-\mu}{k_BT}\Big)\Big]^{-1}.
$$

Here $k_B$ is the Boltzmann constant, $T$ is the plasma temperature,
$\mu$ is the chemical potential of elecrtonic gas.

In an explicit form locally equilibrium function of distribution
can be presented in the form
$$
f^{(0)}(\mathbf{r},\mathbf{p},t)=\Bigg\{1+\exp\Big[\dfrac{\big[\mathbf{p}-
(e/c)\mathbf{A}(\mathbf{r},t)\big]^2}{2mk_BT}+
\dfrac{eU(\mathbf{r},t)-\mu}{k_BT}\Big]\Bigg\}^{-1}.
$$

Let us enter dimensionless electron velocity
$\mathbf{C}(\mathbf{r},t)$,
scalar potential $ \phi(\mathbf{r}, t) $ and chemical potential $ \alpha $

$$
\mathbf{C}(\mathbf{r},\mathbf{p},t)=\dfrac{\mathbf{v}(\mathbf{r},
\mathbf{p},t)}{v_T}=
\dfrac{1}{p_T}\Big[\mathbf{p}-\dfrac{e}{c}\mathbf{A}(\mathbf{r},t)\Big],
$$
$$
\phi(\mathbf{r},t)=\dfrac{eU(\mathbf{r},t)}{k_BT}, \qquad
\alpha=\dfrac{\mu}{k_BT},
$$
where
$v_T=\dfrac{1}{\sqrt{\beta}}$ is the thermal electron velocity,
$\beta=\dfrac{m}{2k_BT}$.

Now locally equilibrium function can be presented by electron velocity
as
$$
f^{(0)}(\mathbf{r},\mathbf{p},t)=
\Big[1+\exp\Big(\dfrac{mv^2(\mathbf{r},\mathbf{p},t)}{2k_BT}+
\dfrac{eU(\mathbf{r},t)-\mu}{k_BT}\Big)
\Big]^{-1},
$$
or, in dimensionless parameters,
$$
f^{(0)}(\mathbf{r},\mathbf{p},t)=\dfrac{1}{1+\exp\big[C^2(\mathbf{r},
\mathbf{p},t)+
\phi(\mathbf{r},t)-\alpha\big]}.
\eqno{(1.11)}
$$

We denote \(\chi=\alpha-\phi\). Then
$$
f^{(0)}=\dfrac{1}{1+e^{C^2-\chi}}.
$$

The quantity $\chi$ it is defined from the law of preservation
of number of particles
$$
\int f d\Omega_F=
\int f^{(0)}d\Omega_F.
$$

Here $d\Omega_F$ is the quantum measure for electrons,
$$
d\Omega_F=\dfrac{2d^3p}{(2\pi\hbar)^3}.
$$

Let us notice, that in case of constant potentials
$U =\const, \mathbf {A} = \const $ the equilibrium
function of distribution (1.11) is the solution of the equation (1.10).

Let us find of electron concentration (numerical density) $N $ and average
electron velocity $ \mathbf {u} $ in the equilibrium condition.
These macroparameters are defined as follows

$$
N(\mathbf{r},t)=\int f(\mathbf{r},\mathbf{p},t)d\Omega_F,
$$
$$
\mathbf{u}(\mathbf{r},t)=\dfrac{1}{N(\mathbf{r},t)}
\int \mathbf{v}(\mathbf{r},\mathbf{p},t)
f(\mathbf{r},\mathbf{p},t)d\Omega_F.
$$

For calculation of these macroparameters in the equilibrium condition
it is necessary to put $f=f^{(0)}$, where $f^{(0)}$ it is defined by equality
(1.11). To designate these macroparameters in an equilibrium condition
let us be through
$N^{(0)}(\mathbf{r},t)$ and $\mathbf{u}^{(0)}(\mathbf{r},t)$.

Let us carry out replacement of the variable of integration
$$
\mathbf{p}-\dfrac{e}{c}\mathbf{A}(\mathbf{r},t)= \mathbf{p}'
$$
in these previous equalities. Then,
passing to integration in spherical coordinates, for
numerical density in an equilibrium condition we receive

$$
N^{(0)}=\dfrac{m^3v_T^3}{\pi^2\hbar^3}f_2(\alpha-\phi),
\eqno{(1.12)}
$$
where
$$
f_2(\alpha-\phi)=\int\limits_{0}^{\infty}\dfrac{x^2\;dx}{1+\exp(x^2+
\phi-\alpha)}=\int\limits_{0}^{\infty}x^2f_F(\alpha-\phi)\,dx.
$$

In the same way, as for numerical density, for average velocity
in an equilibrium condition it is received
$$
\mathbf{u}^{(0)}(\mathbf{r},t)=\dfrac{1}{N^{(0)}}
\int \mathbf{v}(\mathbf{r},\mathbf{p},t)f^{(0)}(\mathbf{r},\mathbf{p},t)
d\Omega_F,
$$
or, in explicit form,
$$
\mathbf{u}^{(0)}(\mathbf{r},t)=\dfrac{2}{N^{(0)}(2\pi\hbar)^3}\int
\dfrac{[\mathbf{p}-(e/c)\mathbf{A}]\;d^3p}
{1+\exp\Big[\dfrac{(\mathbf{p}-(e/c)\mathbf{A})^2}{2k_BTm}+
\dfrac{eU-\mu}{k_Tm}\Big]}.
$$

 After the same replacement of variables
$\mathbf{p}-(e/c)\mathbf{A}(\mathbf{r},t)= \mathbf{p}'$
we receive
$$
\mathbf{u}^{(0)}(\mathbf{r},t)=
\dfrac{2}{N^{(0)}(2\pi\hbar)^3}\int \dfrac{\mathbf{p}'\;d^3p'}
{1+\exp\Big[\dfrac{{p'}^2}{2k_BTm}+\dfrac{eU-\mu}{k_Tm}\Big]}=0.
\eqno{(1.13)}
$$

So, electron velocity in an equilibrium condition agree to
$ (1.13) $ it is equal to zero.

Let us notice, that numerical electron density and their average
velocity satisfy to the usual equation of a continuity
$$
\dfrac{\partial N}{\partial t}+\mathbf{div}(N\mathbf{u})=0.
\eqno{(1.14)}
$$

For  conclusion of the equation of continuity (1.14) it is necessary
integrate the kinetic the equation (1.10) on the quantum measure for
electrons $d\Omega_F $ and use definition of numerical density and average
velocity. Then it is necessary to take advantage of the law of preservation
of number of particles and check up, that integral on a quantum
measure $d\Omega_F $ from Wigner---Vlasov' integral
is equal to zero. Really, we have

$$
\int W[f]\dfrac{2\;d^3p}{(2\pi\hbar)^3}=2\int\int
\Big\{\cdots\Big\}
e^{i\mathbf{p'}\mathbf{a}/\hbar}\delta(\mathbf{a})\,d^3a\,d^3p'=
$$
$$
=2\int \Big\{\cdots\Big\}\Bigg|_{\mathbf{a}=0}d^3p'\equiv 0,
$$
for, as it is easy to check up simple substitution,
$$
 \Big\{\cdots\Big\}\Bigg|_{\mathbf{a}=0}\equiv 0.
$$

Here the symbol $ \{\cdots \} $ means the same expression, as in
the right part of a parity (1.9), i.e.
$$
\{\cdots\}=\left\{
\dfrac{e}{2mc}
\Big[\mathbf{A}(\mathbf{r}+\dfrac{\mathbf{a}}{2},t)+
\mathbf{A}(\mathbf{r}-\dfrac{\mathbf{a}}{2},t)-
2\mathbf{A}(\mathbf{r},t)\Big]\nabla f\right.+
$$\smallskip
$$
+\dfrac{ie}{ mc\hbar}\Big[
\mathbf{A}(\mathbf{r}+\dfrac{\mathbf{a}}{2},t)
-\mathbf{A}(\mathbf{r}-\dfrac{\mathbf{a}}{2},t)\Big]\mathbf{p'}
f+
$$\smallskip
$$
+\dfrac{e}{2mc}\Big[\div{\mathbf{A}(\mathbf{r}+
\dfrac{\mathbf{a}}{2},t)
}+\div{\mathbf{A}(\mathbf{r}-\dfrac{\mathbf{a}}{2},t)}-
2\div{\mathbf{A}(\mathbf{r},t)}\Big]f-
$$\smallskip
$$-
\dfrac{i e^2}{2 mc^2\hbar}\Big[\mathbf{A}^2(\mathbf{r}+
\dfrac{\mathbf{a}}{2},t)-
\mathbf{A}^2(\mathbf{r}-\dfrac{\mathbf{a}}{2},t)\Big]f-
$$\smallskip
$$
-\left. \dfrac{ie}{\hbar}\Big[U(\mathbf{r}+\dfrac{\mathbf{a}}{2},t)-
U(\mathbf{r}-\dfrac{\mathbf{a}}{2},t)\Big]f\right\}
e^{i(\mathbf{p'}-\mathbf{p})\mathbf{a}/\hbar}.
\eqno{(1.9')}
$$

Let us notice, that the left part of the kinetic equation (1.10)
takes standard form for the transport theory  at the following
calibration con\-dition

$$
\div{\mathbf{A}(\mathbf{r},t)}=0.
\eqno{(1.15)}
$$

Thus, i.e. in case of calibration (1.15), the kinetic equation
(1.10) becomes the following form
$$
\dfrac{\partial f}{\partial t}+\mathbf{v}\nabla f=W[f],
\eqno{(1.16)}
$$
in which the Wigner---Vlasov integral is equal

$$
W[f]=
\iint\left\{
\dfrac{e}{2mc}
\Big[\mathbf{A}(\mathbf{r}+\dfrac{\mathbf{a}}{2},t)+
\mathbf{A}(\mathbf{r}-\dfrac{\mathbf{a}}{2},t)-
2\mathbf{A}(\mathbf{r},t)\Big]\nabla f\right.+
$$
\hspace{0.5cm}
$$
+\dfrac{ie}{ mc\hbar}\Big[
\mathbf{A}(\mathbf{r}+\dfrac{\mathbf{a}}{2},t)
-\mathbf{A}(\mathbf{r}-\dfrac{\mathbf{a}}{2},t)\Big]\mathbf{p'}
f-
$$\smallskip
$$
-\dfrac{i e^2}{2 mc^2\hbar}
\Big[\mathbf{A}^2(\mathbf{r}+\dfrac{\mathbf{a}}{2},t)-
\mathbf{A}^2(\mathbf{r}-\dfrac{\mathbf{a}}{2},t)\Big]f-
$$
\hspace{0.5cm}
$$
-\left. \dfrac{ie}{\hbar}\Big[U(\mathbf{r}+\dfrac{\mathbf{a}}{2},t)-
U(\mathbf{r}-\dfrac{\mathbf{a}}{2},t)\Big]f\right\}
e^{i(\mathbf{p'}-\mathbf{p})\mathbf{a}/\hbar}
\dfrac{d^3a\,d^3p'}{(2\pi\hbar)^3}.
\eqno{(1.17)}
$$\medskip

\begin{center}
  {\bf 2. Kinetic equation and its  solution}
\end{center}

Let us consider, that the scalar potential in the equation (1.17)
is equal to zero
$U(\mathbf{r},t)\equiv 0$.

Vector potential we take  as orthogonal to direction of a wave vector
$\mathbf{k}$:
$$
\mathbf{k}\mathbf{A}(\mathbf{r},t)=0.
\eqno{(2.1)}
$$
in the form of the running harmonious wave
$$
\mathbf{A}(\mathbf{r},t)=\mathbf{A}_0
e^{i(\mathbf{k} \mathbf{r}-\omega t)}.
$$

Let us consider the field of vector potential small enough. It
assumption allows to simplify the equation,
not neglecting components, square-law (quadratic) on vector potential
of electromagnetic field.

Then the equation (1.17) becomes simpler
$$
\dfrac{\partial f}{\partial t}+\mathbf{v}\nabla f=W[f].
\eqno{(2.2)}
$$

In the equation (2.2) locally equilibrium distribution Fermi---Dirac'
becomes simpler
$$
f^{(0)}=f^{(0)}(\mathbf{r},\mathbf{p},t)=
\Big[1+\exp\Big(C^2(\mathbf{r},\mathbf{p},t)-\alpha\Big)\Big]^{-1}.
\eqno{(2.3)}
$$

Wigner---Vlasov' integral (1.17) becomes simpler now
also and has the following form

$$
W[f]=\iint\Bigg\{
\dfrac{e}{2mc}
\Big[\mathbf{A}(\mathbf{r}+\dfrac{\mathbf{a}}{2},t)+
\mathbf{A}(\mathbf{r}-\dfrac{\mathbf{a}}{2},t)-
2\mathbf{A}(\mathbf{r},t)\Big]\nabla f(\mathbf{r},\mathbf{p'},t)+
$$
\hspace{0.5cm}
$$
+\dfrac{ie}{mc\hbar }
\Big[\mathbf{A}(\mathbf{r}+\dfrac{\mathbf{a}}{2},t)
-\mathbf{A}(\mathbf{r}-\dfrac{\mathbf{a}}{2},t)\Big]\mathbf{p'}
f(\mathbf{r},\mathbf{p'},t)-
$$
$$
-\dfrac{i e^2}{2 mc^2\hbar}
\Big[\mathbf{A}^2(\mathbf{r}+\dfrac{\mathbf{a}}{2},t)-
\mathbf{A}^2(\mathbf{r}-\dfrac{\mathbf{a}}{2},t)\Big]
f(\mathbf{r},\mathbf{p'},t)\Bigg\}
e^{i(\mathbf{p'}-\mathbf{p})\mathbf{a}/\hbar}
\dfrac{d^3a\,d^3p'}{(2\pi\hbar)^3}.
\eqno{(2.4)}
$$

We note that
$$
\mathbf{A}(\mathbf{r}+\dfrac{\mathbf{a}}{2},t)+
\mathbf{A}(\mathbf{r}-\dfrac{\mathbf{a}}{2},t)-
2\mathbf{A}(\mathbf{r},t)=\mathbf{A}(\mathbf{r},t)
\Big[e^{i\mathbf{k}\mathbf{a}/2}+e^{-i\mathbf{k}\mathbf{a}/2}-2\Big],
$$
$$
\mathbf{A}(\mathbf{r}+\dfrac{\mathbf{a}}{2},t)
-\mathbf{A}(\mathbf{r}-\dfrac{\mathbf{a}}{2},t)=\mathbf{A}(\mathbf{r},t)
\Big[e^{i\mathbf{k}\mathbf{a}/2}-e^{-i\mathbf{k}\mathbf{a}/2}\Big],
$$
$$
\mathbf{A}^2(\mathbf{r}+\dfrac{\mathbf{a}}{2},t)-
\mathbf{A}^2(\mathbf{r}-\dfrac{\mathbf{a}}{2},t)=
\mathbf{A}^2(\mathbf{r},t)
\Big[e^{i\mathbf{k}\mathbf{a}}-e^{-i\mathbf{k}\mathbf{a}}\Big].
$$\smallskip

Substituting these equalities in integral (2.4), we find, that
$$
W[f]=\dfrac{e}{2mc}\mathbf{A}(\mathbf{r},t)\times
$$
$$
\times
\iint \Big[e^{i\mathbf{k}\mathbf{a}/2}+e^{-i\mathbf{k}\mathbf{a}/2}-2\Big]
\nabla f(\mathbf{r},\mathbf{p'},t)
e^{i(\mathbf{p}'-\mathbf{p})\mathbf{a}/\hbar}\dfrac{d^3a\,d^3p'}
{(2\pi\hbar)^3}+
$$
$$
+\dfrac{ie}{mc\hbar}\mathbf{A}(\mathbf{r},t)
\iint \Big[e^{i\mathbf{k}\mathbf{a}/2}-e^{-i\mathbf{k}\mathbf{a}/2}\Big]
\mathbf{p'}f(\mathbf{r},\mathbf{p'},t)
e^{i(\mathbf{p}'-\mathbf{p})\mathbf{a}/\hbar}\dfrac{d^3a\,d^3p'}
{(2\pi\hbar)^3}-
$$
$$
-\dfrac{i e^2}{2 mc^2\hbar}\mathbf{A}^2(\mathbf{r},t)\iint
\Big[e^{i\mathbf{k}\mathbf{a}}-e^{-i\mathbf{k}\mathbf{a}}\Big]
f(\mathbf{r},\mathbf{p'},t)
e^{i(\mathbf{p'}-\mathbf{p})\mathbf{a}/\hbar}
\dfrac{d^3a\,d^3p'}{(2\pi\hbar)^3}.
$$ \vspace{2mm}

Internal integrals in two last summand accordingly are equal

$$
\dfrac{1}{(2\pi\hbar)^3}\int \Big\{\exp\Big(i\Big[\mathbf{p}'-\mathbf{p}+
\frac{\mathbf{k}\hbar}{2}\Big]\dfrac{\mathbf{a}}{\hbar}\Big)+
\exp\Big(i\Big[\mathbf{p}'-\mathbf{p}-
\frac{\mathbf{k}\hbar}{2}\Big]\dfrac{\mathbf{a}}{\hbar}\Big)-
$$
$$
-2\exp\Big(i\Big[\mathbf{p'}-\mathbf{p}\Big]
\frac{\mathbf{k}\hbar}{2}\Big)\Big\}d^3a=
$$
$$
=\delta(\mathbf{p}'-\mathbf{p}+\dfrac{\hbar\mathbf{k}}{2})+
\delta(\mathbf{p}'-\mathbf{p}-\dfrac{\hbar\mathbf{k}}{2})-2
\delta(\mathbf{p'}-\mathbf{p}),
$$ \medskip
$$
\dfrac{1}{(2\pi\hbar)^3}\int \Big\{\exp\Big(i\Big[\mathbf{p}'-\mathbf{p}+
\frac{\mathbf{k}\hbar}{2}\Big]\dfrac{\mathbf{a}}{\hbar}\Big)-
\exp\Big(i\Big[\mathbf{p}'-\mathbf{p}-
\frac{\mathbf{k}\hbar}{2}\Big]\dfrac{\mathbf{a}}{\hbar}\Big)\Big\}d^3a=
$$
$$
=\delta(\mathbf{p}'-\mathbf{p}+\dfrac{\hbar\mathbf{k}}{2})-
\delta(\mathbf{p}'-\mathbf{p}-\dfrac{\hbar\mathbf{k}}{2}),
$$\medskip
and

$$
\dfrac{1}{(2\pi\hbar)^3}\int \bigg\{\exp\Big(i\Big[\mathbf{p}'-\mathbf{p}+
{\mathbf{k}\hbar}\Big]\dfrac{\mathbf{a}}{\hbar}\Big)-
\exp\Big(i\Big[\mathbf{p}'-\mathbf{p}-
{\mathbf{k}\hbar}\Big]\dfrac{\mathbf{a}}{\hbar}\Big)\bigg\}d^3a=
$$
$$
=\delta(\mathbf{p}'-\mathbf{p}+{\hbar\mathbf{k}})-
\delta(\mathbf{p}'-\mathbf{p}-{\hbar\mathbf{k}}).
$$\vspace{2mm}

Now the Wigner---Vlasov integral is expressed through one-dimen\-sio\-nal integral
$$
W[f]=\mathbf{A}(\mathbf{r},t)\dfrac{e}{2mc}\times $$$$ \times
\int \Big[\delta(\mathbf{p}'-\mathbf{p}+\dfrac{\hbar\mathbf{k}}{2})+
\delta(\mathbf{p}'-\mathbf{p}-\dfrac{\hbar\mathbf{k}}{2})-
2\delta(\mathbf{p'}-\mathbf{p})\Big]
\nabla f(\mathbf{r},\mathbf{p}',t)\,d^3p'+
$$ \medskip
$$
+\mathbf{A}(\mathbf{r},t)\dfrac{ie}{mc\hbar}
\int \Big[\delta(\mathbf{p}'-\mathbf{p}+\dfrac{\hbar\mathbf{k}}{2})-
\delta(\mathbf{p}'-\mathbf{p}-\dfrac{\hbar\mathbf{k}}{2})\Big]
\mathbf{p}'f(\mathbf{r},\mathbf{p}',t)\,d^3p' -
$$\vspace{0.3cm}
$$
-\mathbf{A}^2(\mathbf{r},t)\dfrac{i e^2}{2 mc^2\hbar}\int
\Big[\delta(\mathbf{p}'-\mathbf{p}+{\hbar\mathbf{k}})-
\delta(\mathbf{p}'-\mathbf{p}-{\hbar\mathbf{k}})\Big]
f(\mathbf{r},\mathbf{p}',t)\,d^3p'.
$$\medskip

It is necessary to take advantage of properties Dirac delta--function.
We receive, that
$$
W[f]=\mathbf{A}(\mathbf{r},t)\dfrac{e}{2mc}
\Big[\nabla f(\mathbf{r},\mathbf{p}-\dfrac{\hbar \mathbf{k}}{2},t)+
\nabla f(\mathbf{r},\mathbf{p}+\dfrac{\hbar \mathbf{k}}{2},t)-
2\nabla f(\mathbf{r},\mathbf{p},t)\Big]+
$$\vspace{0.3cm}
$$
+\mathbf{A}(\mathbf{r},t)\dfrac{ie}{mc\hbar}
\Big[\Big(\mathbf{p}-\dfrac{\hbar \mathbf{k}}{2}\Big)f(\mathbf{r},
\mathbf{p}-\dfrac{\hbar \mathbf{k}}{2},t)-
\Big(\mathbf{p}+\dfrac{\hbar \mathbf{k}}{2}\Big)f(\mathbf{r},
\mathbf{p}+\dfrac{\hbar \mathbf{k}}{2},t)\Big]-
$$\vspace{0.3cm}
$$
-\mathbf{A}^2(\mathbf{r},t)\dfrac{i e^2}{2 mc^2\hbar}
\Big[f(\mathbf{r},\mathbf{p}-{\hbar \mathbf{k}},t)-
f(\mathbf{r},\mathbf{p}+{\hbar \mathbf{k}},t)\Big].
$$\medskip

Let us transform the received expression of Wigner---Vlasov
integral
$$
W[f]=\mathbf{A}(\mathbf{r},t)\dfrac{e}{2mc}
\Big[\nabla f(\mathbf{r},\mathbf{p}-\dfrac{\hbar \mathbf{k}}{2},t)+
\nabla f(\mathbf{r},\mathbf{p}+\dfrac{\hbar \mathbf{k}}{2},t)-
2\nabla f(\mathbf{r},\mathbf{p},t)\Big]+
$$\vspace{0.3cm}
$$
+\mathbf{A}(\mathbf{r},t)\dfrac{ie}{mc\hbar}
\Big\{\mathbf{p}\Big[f(\mathbf{r},
\mathbf{p}-\dfrac{\hbar \mathbf{k}}{2},t)-f(\mathbf{r},
\mathbf{p}+\dfrac{\hbar \mathbf{k}}{2},t)\Big]-
$$
\vspace{0.3cm}
$$
-\dfrac{\hbar\mathbf{k}}{2}\Big[f(\mathbf{r},
\mathbf{p}-\dfrac{\hbar \mathbf{k}}{2},t)+f(\mathbf{r},
\mathbf{p}+\dfrac{\hbar \mathbf{k}}{2},t)\Big]\Big\}-
$$\vspace{0.3cm}
$$
-\mathbf{A}^2(\mathbf{r},t)\dfrac{i e^2}{2 mc^2\hbar}
\Big[f(\mathbf{r},\mathbf{p}-{\hbar \mathbf{k}},t)-
f(\mathbf{r},\mathbf{p}+{\hbar \mathbf{k}},t)\Big].
$$

Let us enter following designations
$$
f_{\pm}\equiv f(\mathbf{r},\mathbf{p}\mp\dfrac{\hbar
\mathbf{k}}{2},t), \quad
f_{++}\equiv f(\mathbf{r},\mathbf{p}-{\hbar\mathbf{k}},t),\quad
f_{--}\equiv f(\mathbf{r},\mathbf{p}+{\hbar\mathbf{k}},t).
$$

Then Wigner---Vlasov integral (2.4) can be rewritten more shortly
$$
W[f]=\mathbf{A}(\mathbf{r},t)\dfrac{e}{2mc}\Big(\nabla f_++ \nabla f_-
-2\nabla f\Big)+
$$
$$
+\mathbf{A}(\mathbf{r},t)\dfrac{ie}{mc\hbar}\mathbf{p}\Big(f_+-f_-\Big)
-\mathbf{A}^2(\mathbf{r},t)\dfrac{i e^2}{2 mc^2\hbar}
\Big(f_{++}-f_{--}\Big).
\eqno{(2.5)}
$$ \medskip

Further it is more convenient to use dimensionless velocity
$ \mathbf {C} $ in the form
$$
\mathbf{C}=\dfrac{\mathbf{v}}{v_T}=\dfrac{\mathbf{p}}{p_T}-
\dfrac{e}{cp_T}\mathbf{A}(\mathbf{r},t)\equiv \mathbf{P}-
\dfrac{e}{cp_T}\mathbf{A}(\mathbf{r},t),
$$
where $\mathbf{P}=\dfrac{\mathbf{p}}{p_T}$ is the dimensionless momentum.

Let here and more low expression $ \mathbf {P} \mathbf {A} $ means
scalar product.

In quadratic (square-law) approximation on vector potential
$\mathbf{A}(\mathbf{r},t) $
Wigner's  function $f $ in the first summand in Wigner---Vlasov
integral it is necessary to replace on locally equilibrium
Fermi---Dirac's distribution (2.3) $f^{(0)}(\mathbf{r,P},t)$,
i.e. we consider that $f=f^{(0)}(\mathbf{r,P},t)$, where
$$
f^{(0)}=f^{(0)}(\mathbf{r},\mathbf{P},t)=
\Big[1+\exp\Big(C^2(\mathbf{r},\mathbf{P},t)-\alpha\Big)\Big]^{-1}.
$$

In the third summand it is necessary to replace Wigner's  function $f $ on
absolute Fermian, i.e. we consider that $f=f_F (P) $, where

$$
f_F(P)=\dfrac{1}{1+\exp (P^2-\alpha)}, \qquad \alpha=\const.
$$

In the second summand it is necessary to replace Wigner's  function $f $ on
its linear approximation found in our work \cite{Lat1} i.e. we consider that

$$
f=f^{(0)}-\mathbf{PA}(\mathbf{r},t)\Big[\dfrac{2e}{cp_T}g(P)+
\dfrac{ev_T}{c\hbar}\dfrac{f_F^+-f_F^-}{\omega-v_T\mathbf{kP}}\Big],
$$
where
$$
g(P)=\dfrac{e^{P^2-\alpha}}{(1+e^{P^2-\alpha})^2}=
-\dfrac{\partial}{\partial P^2}f_F(P).
$$

Let us notice, that in linear approximation
$$
f^{(0)}=f_F(P)+g(P)\dfrac{2e}{cp_T}\mathbf{PA}(\mathbf{r},t).
$$
Hence, Wigner's  function $f $ is represented in the form
$$
f=f_F(P)-\dfrac{ev_T}{c\hbar}\mathbf{PA}(\mathbf{r},t)\dfrac{f_F^+-f_F^-}
{\omega-v_T\mathbf{kP}}.
$$

Thus the Wigner---Vlasov integral (2.5) will look like
$$
W[f]=\dfrac{e}{2mc}\mathbf{A}(\mathbf{r},t)\Big[\nabla f^{(0)+}+
\nabla f^{(0)-}-2\nabla f^{(0)}\Big]+
$$
$$
+\dfrac{iev_T}{c\hbar}\mathbf{P}\mathbf{A}
(\mathbf{r},t)\Big[f_F^+-f_F^--\dfrac{ev_T}{c\hbar}\mathbf{PA}(\mathbf{r},t)
\dfrac{(f_F^+-f_F^-)^+-(f_F^+-f_F^-)^-}{\omega-v_T\mathbf{kP}}\Big]-
$$
$$
-\mathbf{A}^2(\mathbf{r},t)\dfrac{i e^2}{2 mc^2\hbar}
\Big(f_F^{++}-f_F^{--}\Big).
\eqno{(2.6)}
$$\medskip

Here

$$
f^{(0)\pm}=f^{(0)\pm}(\mathbf{r,P},t)=
\dfrac{1}{1+\exp\Big[\Big(\mathbf{P}\mp \dfrac{\hbar \mathbf{k}}{2p_T}-
\dfrac{e}{cp_T}\mathbf{A}(\mathbf{r},t)\Big)^2-\alpha\Big]},
$$

$$
f^{(0)}=f^{(0)}(\mathbf{r,P},t)=
\dfrac{1}{1+\exp\Big[\Big(\mathbf{P}-
\dfrac{e}{cp_T}\mathbf{A}(\mathbf{r},t)\Big)^2-\alpha\Big]},
$$

$$
f_F^{\pm}\equiv f_F^{\pm}(\mathbf{P})=
\dfrac{1}{1+
\exp\Big[\Big(\mathbf{P}\mp \dfrac{\hbar \mathbf{k}}{2p_T}\Big)^2
-\alpha\Big]},
$$
$$
f_F^{++}\equiv f_F^{++}(\mathbf{P})=
\dfrac{1}{1+
\exp\Big[\Big(\mathbf{P}-\dfrac{\hbar \mathbf{k}}{p_T}\Big)^2
-\alpha\Big]},
$$
$$
f_F^{--}\equiv f_F^{--}(\mathbf{P})=
\dfrac{1}{1+
\exp\Big[\Big(\mathbf{P}+\dfrac{\hbar \mathbf{k}}{p_T}\Big)^2
-\alpha\Big]},
$$
and $p_T=mv_T$ is the thermal electron momentum.

It is possible to present these designations more shortly

$$
f^{(0)\pm}=f^{(0)}(\mathbf{r},C_{\pm},t)=\dfrac{1}{1+e^{C_{\pm}^2-\alpha}},
$$

$$
f_F^{\pm}\equiv f_F(P_{\pm})=\dfrac{1}{1+e^{P^2_{\pm}-\alpha}},
$$
$$
f_F^{++}\equiv f_F(P_{++})=\dfrac{1}{1+e^{P^2_{++}-\alpha}},
$$
$$
f_F^{--}\equiv f_F(P_{--})=\dfrac{1}{1+e^{P^2_{--}-\alpha}}.
$$

Here

$$
{C}_{\pm}^2=\Big(\mathbf{P}\mp\dfrac{\hbar \mathbf{k}}{2p_T}-
\dfrac{e}{cp_T}\mathbf{A}(\mathbf{r},t)\Big)^2,
$$\medskip
$$
P^2_{\pm}=\Big(\mathbf{P}\mp\dfrac{\hbar \mathbf{k}}{2p_T}\Big)^2
=\Big(P_x\mp\dfrac{\hbar k_x}{2p_T}\Big)^2+
\Big(P_y\mp\dfrac{\hbar k_y}{2p_T}\Big)^2+
\Big(P_z\mp\dfrac{\hbar k_z}{2p_T}\Big)^2,
$$
or
$$
P^2_{\pm}=\dfrac{\Big(p_x\mp\dfrac{\hbar k_x}{2}\Big)^2+
\Big(p_y\mp\dfrac{\hbar k_y}{2}\Big)^2+
\Big(p_z\mp\dfrac{\hbar k_z}{2}\Big)^2}{p_T^2}.
$$
Besides,
$$
P^2_{++}=\Big(\mathbf{P}-\dfrac{\hbar \mathbf{k}}{p_T}\Big)^2
=\Big(P_x-\dfrac{\hbar k_x}{p_T}\Big)^2+
\Big(P_y-\dfrac{\hbar k_y}{p_T}\Big)^2+
\Big(P_z-\dfrac{\hbar k_z}{p_T}\Big)^2,
$$
or
$$
P^2_{++}=\dfrac{\Big(p_x-{\hbar k_x}\Big)^2+
\Big(p_y-{\hbar k_y}\Big)^2+\Big(p_z-{\hbar k_z}\Big)^2}{p_T^2}.
$$
Analogous,
$$
P^2_{--}=\Big(\mathbf{P}+\dfrac{\hbar \mathbf{k}}{p_T}\Big)^2
=\Big(P_x+\dfrac{\hbar k_x}{p_T}\Big)^2+
\Big(P_y+\dfrac{\hbar k_y}{p_T}\Big)^2+
\Big(P_z+\dfrac{\hbar k_z}{p_T}\Big)^2,
$$
or
$$
P^2_{--}=\dfrac{\Big(p_x+{\hbar k_x}\Big)^2+
\Big(p_y+{\hbar k_y}\Big)^2+\Big(p_z+{\hbar k_z}\Big)^2}{p_T^2}.
$$

Let us show, that the first summand in Wigner---Vlasov integral
(2.6) equally to zero. According to  statement  problem vector potential
of electromagnetic field varies along an axis $x $. Hence,
gradient of locally Fermi---Dirac' equilibrium distribution
is proportional to a vector $\mathbf{k}$:
$\nabla f^{(0)\pm}\sim \mathbf{k}$, $\nabla f^{(0)}\sim \mathbf{k}$.
Therefore
$$
\mathbf{A}(\mathbf{r},t)\Big[\nabla f^{(0)+}+
\nabla f^{(0)-}-2\nabla f^{(0)}\Big]\sim\mathbf{Ak}=0.
$$

Thus, the Wigner---Vlasov integral is equal
$$
W[f]=$$$$+\dfrac{iev_T}{c\hbar}\mathbf{P}\mathbf{A}
(\mathbf{r},t)\Big[f_F^+-f_F^--\dfrac{ev_T}{c\hbar}\mathbf{PA}(\mathbf{r},t)
\dfrac{(f_F^+-f_F^-)^+-(f_F^+-f_F^-)^-}{\omega-v_T\mathbf{kP}}\Big]-
$$
$$
-\mathbf{A}^2(\mathbf{r},t)\dfrac{i e^2}{2 mc^2\hbar}
\Big(f_F^{++}-f_F^{--}\Big).
\eqno{(2.7)}
$$\medskip

We note, that
$$
f_F^{+-}=f_F^{-+}=f_F=f_F(P).
$$
Therefore, the Wigner---Vlasov integral (2.7) is equal
$$
W[f]=\dfrac{iev_T}{c\hbar}\mathbf{P}\mathbf{A}
(\mathbf{r},t)\Big[f_F^+-f_F^--\dfrac{ev_T}{c\hbar}\mathbf{PA}(\mathbf{r},t)
\dfrac{f_F^{++}+f_F^{--}-2f_F}{\omega-v_T\mathbf{kP}}\Big]-
$$
$$
-\mathbf{A}^2(\mathbf{r},t)\dfrac{i e^2}{2 mc^2\hbar}
\Big(f_F^{++}-f_F^{--}\Big).
\eqno{(2.8)}
$$\medskip

Let us return to the solution of the equation (2.2).
We will search for Wigner' function in the form, quadratic (square-law)
concerning vector potential
$\mathbf{A}(\mathbf{r},t)$:

$$
f=f^{(0)}-\mathbf{PA}(\mathbf{r},t)\Big[\dfrac{2e}{cp_T}g(P)+
\dfrac{ev_T}{c\hbar}\dfrac{f_F^+-f_F^-}{\omega-v_T\mathbf{kP}}\Big]+
[\mathbf{A}(\mathbf{r},t)]^2
h(\mathbf{P})=
$$ \vspace{2mm}
$$
=f_F(P)-\dfrac{ev_T}{c\hbar}\mathbf{PA}(\mathbf{r},t)
\dfrac{f_F^+-f_F^-}{\omega-v_T\mathbf{kP}}+[\mathbf{A}(\mathbf{r},t)]^2
h(\mathbf{P}),
$$ \vspace{3mm}
where $h(\mathbf{P})$ is the new unknown function.

We receive equation, from which we find
$$
[\mathbf{A}(\mathbf{r},t)]^2h(\mathbf{P})=
\dfrac{(ev_T)^2}{2(c\hbar)^2}[\mathbf{PA}(\mathbf{r},t)]^2
\dfrac{f_F^{++}+f_F^{--}-2f_F}{(\omega-v_T\mathbf{kP})^2}+
$$ \vspace{2mm}
$$
+\dfrac{e^2}{4mc^2\hbar}[\mathbf{A}(\mathbf{r},t)]^2\dfrac{f_F^{++}-
f_F^{--}}{\omega-v_T\mathbf{kP}}.
$$\vspace{2mm}

By means of last two equalities let us construct Wigner' function
in the second approximation on the vector field
$\mathbf{A}(\mathbf{r},t)$:\medskip
$$
f=f^{(0)}-\mathbf{PA}(\mathbf{r},t)\Big[\dfrac{2e}{cp_T}g(P)+
\dfrac{ev_T}{c\hbar}\dfrac{f_F^+-f_F^-}{\omega-v_T\mathbf{kP}}\Big]+
$$\medskip
$$
+\dfrac{(ev_T)^2}{2(c\hbar)^2}[\mathbf{PA}(\mathbf{r},t)]^2
\dfrac{f_F^{++}+f_F^{--}-2f_F}{(\omega-v_T\mathbf{kP})^2}+
$$\medskip
$$
+\dfrac{e^2}{4mc^2\hbar}[\mathbf{A}(\mathbf{r},t)]^2\dfrac{f_F^{++}-
f_F^{--}}{\omega-v_T\mathbf{kP}}.
\eqno{(2.9)}
$$ \medskip

This function represents quadratic (square-law) decomposition of
dist\-ri\-bu\-tion function on vector potential
$\mathbf{A}(\mathbf{r},t)$.

\begin{center}
\bf  3. Density of electric current in quantum plasmas
\end{center}

By definition, the density of electric current is equal
$$
\mathbf{j}(\mathbf{r},t)=e\int \mathbf{v}(\mathbf{r},\mathbf{p},t)
f(\mathbf{r},\mathbf{p},t)\dfrac{2\,d^3p}{(2\pi\hbar)^3}.
\eqno{(3.1)}
$$

In our work \cite{Lat1} it is shown, that density of
electric current in an equilibrium condition (the calibrating current)
is equal to zero
$$
\mathbf{j^{(0)}}(\mathbf{r},t)=e\int \mathbf{v}(\mathbf{r},\mathbf{p},t)
f^{(0)}(\mathbf{r},\mathbf{p},t)\dfrac{2\,d^3p}{(2\pi\hbar)^3}=0.
$$

Hence, for the density of electric current we  receive the following
expression
$$
\mathbf{j}(\mathbf{r},t)=e\int \mathbf{v}(\mathbf{r},\mathbf{p},t)
\Bigg[-\mathbf{PA}(\mathbf{r},t)\Big[\dfrac{2e}{cp_T}g(P)+
\dfrac{ev_T}{c\hbar}\dfrac{f_F^+-f_F^-}{\omega-v_T\mathbf{kP}}\Big]+
$$
$$
+\dfrac{(ev_T)^2}{2(c\hbar)^2}[\mathbf{PA}(\mathbf{r},t)]^2
\dfrac{f_F^{++}+f_F^{--}-2f_F}{(\omega-v_T\mathbf{kP})^2}+
$$
$$
+\dfrac{e^2}{4mc^2\hbar}[\mathbf{A}(\mathbf{r},t)]^2\dfrac{f_F^{++}-
f_F^{--}}{\omega-v_T\mathbf{kP}} \Bigg]
\dfrac{2\,d^3p}{(2\pi\hbar)^3}.
$$

Substituting in equality (3.1) explicit expression for velocity
$$
\mathbf{v}(\mathbf{r},\mathbf{P},t)=
\dfrac{\mathbf{p}}{m}-\dfrac{e \mathbf{A}(\mathbf{r},t)}{mc}=
\dfrac{p_T\mathbf{P}}{m}-\dfrac{e \mathbf{A}(\mathbf{r},t)}{mc}=
v_T\mathbf{P}-\dfrac{e \mathbf{A}(\mathbf{r},t)}{mc}.
$$
and, leaving linear and quadratic (square-law) expressions concerning
vector potential of the field, we receive \medskip
$$
\mathbf{j}(\mathbf{r},t)=-
\dfrac{2e^2p_T^4}{(2\pi \hbar)^3mc}\int
\mathbf{P}\big[\mathbf{P}\mathbf{A}\big]\Big[
\dfrac{2}{p_T}g(P)+\dfrac{v_T}{\hbar}\dfrac{f_F^+-f_F^-}{\omega
-v_T\mathbf{kP}}\Big]d^3P+
$$ \medskip
$$
+\dfrac{2e^3p_T^3}{(2\pi \hbar)^3mc^2}\mathbf{A}
\int\big[\mathbf{P}\mathbf{A}\big]
\Big[
\dfrac{2}{p_T}g(P)+\dfrac{v_T}{\hbar}\dfrac{f_F^+-f_F^-}{\omega
-v_T\mathbf{kP}}\Big]d^3P+
$$ \medskip
$$
+\dfrac{2e^3p_T^4}{(2\pi\hbar)^3mc^2\hbar}
\int\mathbf{P} \Bigg[\dfrac{mv_T^2}{2\hbar}[\mathbf{PA}]^2
\dfrac{f_F^{++}+f_F^{--}-2f_F}
{(\omega-v_T\mathbf{kP})^2}+\dfrac{\mathbf{A}^2}{4}
\dfrac{f_F^{++}-f_F^{--}}{\omega-v_T\mathbf{kP}}\Bigg]d^3P.
\eqno{(3.2)}
$$ \medskip

The first summand in (3.2) is linear expression
of the density of electric current, found, in particular, in our previous
work \cite{Lat1}. Other summands are
the quadratic (square-law) amendments caused by vector potential
of electromagnetic field.

\begin{center}
\bf  4. Linear part of density of electric current
\end{center}

Linear part of density of electric current
$$
\mathbf{j}_{\rm linear}(\mathbf{r},t)=$$$$=-
\dfrac{2e^2p_T^4}{(2\pi \hbar)^3mc}\int
\mathbf{P}\big[\mathbf{P}\mathbf{A}\big]\Big[
\dfrac{2}{p_T}g(P)+\dfrac{v_T}{\hbar}\dfrac{f_F^+-f_F^-}{\omega
-v_T\mathbf{kP}}\Big]d^3P.
\eqno{(4.1)}
$$
we represent in invariant form.

We note, that
$$
\Big(\mathbf{P}\mp \dfrac{\hbar \mathbf{k}}{2p_T}\Big)^2=
P^2\mp
\dfrac{\hbar}{p_T}\mathbf{Pk}+\Big(\dfrac{\hbar}{2p_T}\Big)^2k^2.
$$

We take the unity vector $ \mathbf {e}_1 =\dfrac{\mathbf {A}}{A} $,
directed along the vector $ \mathbf {A} $. Then equality (4.1)
it is possible to write down in the form
$$
\mathbf{j}_{\rm linear}(\mathbf{r},t)=-\dfrac{2e^2p_T^4 A(\mathbf{r},t)}
{(2\pi \hbar)^3mc}
\int \big(\mathbf{P}\mathbf{e}_1\big)
\big[\mathbf{P}\mathbf{A}(\mathbf{r},t)\big]\mathbf{P}\times
$$
$$\times\Big[
\dfrac{2}{p_T}g(P)+\dfrac{v_T}{\hbar}\dfrac{f_F^+-f_F^-}{\omega
-v_T\mathbf{kP}}\Big]\,d^3P.
\eqno{(4.2)}
$$

Take into account symmetry  value of the first integral (4.2) will not
change, if vector $ \mathbf{e}_1$ to replace with any other unity vector
$\mathbf{e}_2$, perpendicular to the vector $\mathbf{k}$, i.e.

$$
\qquad \qquad \mathbf{e}_2=\dfrac{\mathbf{A} \times \mathbf{k}}
{|\mathbf{A} \times \mathbf{k}|}=
\dfrac{\mathbf{A} \times \mathbf{k}}{Ak},
$$
where $\mathbf{A} \times \mathbf{k}$ is the vector product.

Let us develop the vector $ \mathbf{P} $ by three orthogonal directions
$\mathbf{e}_1$, $\mathbf{e}_2$ и
$\mathbf{n}=\dfrac{\mathbf{k}}{k}$:
$$
\mathbf{P}=(\mathbf{Pn})\mathbf{n}+(\mathbf{P}\mathbf{e}_1)
\mathbf{e}_1+(\mathbf{P}\mathbf{e}_2)\mathbf{e}_2.
$$

By means of this decomposition it is received
$$
(\mathbf{PA})\mathbf{P}=A(\mathbf{P}\mathbf{e}_1)\mathbf{P}=
$$
$$
=A(\mathbf{P}\mathbf{e}_1)(\mathbf{Pn})\mathbf{n}+
A(\mathbf{P}\mathbf{e}_1)^2\mathbf{e}_1+A(\mathbf{P}\mathbf{e}_1)
(\mathbf{P}\mathbf{e}_2)\mathbf{e}_2.
$$

Substituting this decomposition in (4.2), and, considering, that integ\-rals
on odd functions on the symmetric interval are equal to zero, we receive
$$
\mathbf{j}_{\rm linear}(\mathbf{r},t)=-\dfrac{2e^2p_T^4 \mathbf{A}(\mathbf{r},t)}
{(2\pi\hbar)^3mc}\int (\mathbf{P}\mathbf{e}_1)^2
\Big[
\dfrac{2}{p_T}g(P)+\dfrac{v_T}{\hbar}\dfrac{f_F^+-f_F^-}{\omega
-v_T\mathbf{kP}}\Big].
\eqno{(4.3)}
$$

In view of symmetry quantity of integral will not change,
if the vector $ \mathbf {e}_1$ to replace with any other unity
vector $ \mathbf {e}_2$, which is perpendicular to the vector $ \mathbf{k} $.
Therefore
$$
\int \big(\mathbf{e}_1\mathbf{P}\big)^2[\cdots]d^3P=
\int \big(\mathbf{e}_2\mathbf{P}\big)^2[\cdots]d^3P=
$$
$$
=\dfrac{1}{2}\int \Big[\Big(\mathbf{e}_1 \mathbf{P}\Big)^2
+\Big(\mathbf{e}_2\mathbf{P}\Big)^2\Big]
[\cdots]d^3P.
$$

Let us notice, that the square of length of a vector $ \mathbf{P} $ is equal
$$
P^2=(\mathbf{P}\mathbf{e}_1)^2+(\mathbf{P}\mathbf{e}_2)^2+
(\mathbf{Pn})^2,
$$
whence
$$
\big(\mathbf{e}_1\mathbf{P}\big)^2
+\big(\mathbf{e}_2\mathbf{P}\big)^2=P^2-
\dfrac{(\mathbf{P}\mathbf{k})^2}{k^2}=P^2-(\mathbf{Pn})^2=
P_\perp^2,
$$
where $P_\perp$ is the projection of vector
$\mathbf{P}$ on a straight line, which is
perpendicular to planes $ (\mathbf {e}_1, \mathbf {e}_2) $.

From here for current density it is received the following expression
$$
\mathbf{j}_{\rm linear}(\mathbf{r},t)=-
\dfrac{e^2p_T^4\mathbf{A}(\mathbf{r},t)}{(2\pi \hbar)^3mc}\int
\Bigg[
\dfrac{2}{p_T}g(P)+
\dfrac{v_T}{\hbar}\dfrac{f_F^+(\mathbf{P})-f_F^-(\mathbf{P})}
{\omega-v_T\mathbf{kP}}\Bigg]P_\perp^2 d^3P.
$$

\begin{center}
  \bf 5. Desity of longitudinal electric current
\end{center}

Let us show, that vector potential of the electromagnetic field besides
transversal current generates also a longitudinal electric current.
From the formula (3.2) it is visible, that the density of
longitudinal current is defined last summand.

Let us return to expression (3.2), square-law on the vector
potential of the electromagnetic field.

Vector potential of the field we will direct along an axis $y $
$$
\mathbf{A}(\mathbf{r},t)=A(\mathbf{r},t)\mathbf{e}_y,
$$
and we direct wave vector $\mathbf{k}$ along an axis $x$:
$$
\mathbf{k}=k\mathbf{e}_x.
$$

Then
$$
\mathbf{A}(\mathbf{r},t)=\mathbf{A}(x,t)=\{0, A_ye^{i(kx-\omega t)},0\},\qquad
\mathbf{kP}=kP_x,
$$
$$
\mathbf{PA}(\mathbf{r},t)=A(\mathbf{r},t)P_y,  \qquad
\mathbf{A}(\mathbf{PA})=A^2(\mathbf{r},t)P_y\mathbf{e}_y,
$$
$$
[\mathbf{PA}]^2=A^2(\mathbf{r},t)P_y^2, \qquad \mathbf{A}^2=A^2(\mathbf{r},t).
$$

According to (3.2) the density of longitudinal current equals
$$
\mathbf{j}_{\rm long}(\mathbf{r},t)=
\dfrac{e^3p_T^4A^2(\mathbf{r},t)}{2(2\pi\hbar)^3mc^2\hbar}\times
$$
$$
\times\int\mathbf{P} \Bigg[2\dfrac{mv_T^2}{\hbar}P_y^2
\dfrac{f_F^{++}+f_F^{--}-2f_F}
{(\omega-v_TkP_x)^2}+
\dfrac{f_F^{++}-f_F^{--}}{\omega-v_T{kP_x}}\Bigg]d^3P.
\eqno{(5.1)}
$$

This vector has only  the one nonzero  first  component,
therefore we can write down, that

$$
\mathbf{j}_{\rm long}(\mathbf{r},t)=j_{x,\rm long}(x,t)\{1,0,0\},
$$
where the quantity of current density is defined by expression
$$
j_{x}^{\rm long}(x,t)=
\dfrac{e^3p_T^4A^2(\mathbf{r},t)}{2(2\pi\hbar)^3mc^2\hbar}\times
$$
$$
\times \Bigg[2\dfrac{mv_T^2}{\hbar}\int
\dfrac{f_F^{++}+f_F^{--}-2f_F}{(\omega-v_TkP_x)^2}P_xP_y^2d^3P+
\int\dfrac{f_F^{++}-f_F^{--}}{\omega-v_T{kP_x}}P_xd^3P\Bigg].
\eqno{(5.2)}
$$

Let us consider the first integral from (5.2).
We will calculate the internal
integrals in a plane $ (P_y, P_z) $, passing to polar coordinates
$$
\int f_F^{++}P_y^2dP_ydP_z=\int\limits_{0}^{2\pi}\int\limits_{0}^{\infty}
\dfrac{\cos^2\varphi \rho^3 d\varphi d\rho}{1+\exp\Big((P_x-\dfrac{\hbar k}
{p_T})^2+\rho^2-\alpha\Big)}=
$$
$$
=\pi\int\limits_{0}^{\infty}\dfrac{\rho^3d\rho}{1+\exp\Big((P_x-\dfrac{\hbar k}
{p_T})^2+\rho^2-\alpha\Big)}=
$$
$$
=\pi \int\limits_{0}^{\infty}
\rho\ln(1+\exp\Big(-(P_x-\dfrac{\hbar k}
{p_T})^2-\rho^2+\alpha\Big))d\rho.
$$

Similarly, other integrals are equal
$$
\int f_F^{--}P_y^2dP_ydP_z=\int\limits_{0}^{2\pi}\int\limits_{0}^{\infty}
\dfrac{\cos^2\varphi \rho^3 d\varphi d\rho}{1+\exp\Big((P_x+\dfrac{\hbar k}
{p_T})^2+\rho^2-\alpha\Big)}=
$$
$$
=\pi \int\limits_{0}^{\infty}
\rho\ln(1+\exp\Big(-(P_x+\dfrac{\hbar k}
{p_T})^2-\rho^2+\alpha\Big))d\rho,
$$
and
$$
\int f_FP_y^2dP_ydP_z=\int\limits_{0}^{2\pi}\int\limits_{0}^{\infty}
\dfrac{\cos^2\varphi \rho^3 d\varphi d\rho}{1+\exp\Big(P_x^2+\rho^2-
\alpha\Big)}=
$$
$$
=\pi \int\limits_{0}^{\infty}
\rho\ln(1+\exp(-P_x^2-\rho^2+\alpha))d\rho.
$$

We enter designations
$$
l_1^{++}(P_x,\alpha)=\int\limits_{0}^{\infty}\rho\ln(1+
e^{-(P_x-\frac{\hbar k}{p_T})^2-\rho^2+\alpha})d\rho,
$$
$$
l_1^{--}(P_x,\alpha)=\int\limits_{0}^{\infty}\rho\ln(1+
e^{-(P_x+\frac{\hbar k}{p_T})^2-\rho^2+\alpha})d\rho,
$$
$$
l_1(P_x,\alpha)=\int\limits_{0}^{\infty}\rho\ln(1+
e^{-P_x^2-\rho^2+\alpha})d\rho.
$$

Hence, the first integral from (5.2) is equal
$$
\int
\dfrac{f_F^{++}+f_F^{--}-2f_F}{(\omega-v_TkP_x)^2}P_xP_y^2d^3P=
$$
$$
=\pi \int\limits_{-\infty}^{\infty}\dfrac{l_1^{++}(\tau,\alpha)+
l_1^{--}(\tau,\alpha)-2l_1(\tau,\alpha)}{(\omega-v_Tk\tau )^2}\tau d\tau.
$$

We consider the second integral from (5.2)
$$
\int\dfrac{f_F^{++}-f_F^{--}}{\omega-v_T{kP_x}}P_xd^3P=
$$
$$
=\int\limits_{-\infty}^{\infty}\dfrac{P_xdP_x}{\omega-v_TkP_x}
\int\limits_{-\infty}^{\infty}\int\limits_{-\infty}^{\infty}
\dfrac{dP_ydP_z}{1+\exp\Big(\Big(P_x-\dfrac{\hbar k}{p_T}\Big)^2+
P_y^2+P_z^2-\alpha\Big)}-
$$
$$
-\int\limits_{-\infty}^{\infty}\dfrac{P_xdP_x}{\omega-v_TkP_x}
\int\limits_{-\infty}^{\infty}\int\limits_{-\infty}^{\infty}
\dfrac{dP_ydP_z}{1+\exp\Big(\Big(P_x+\dfrac{\hbar k}{p_T}\Big)^2+
P_y^2+P_z^2-\alpha\Big)}.
$$

Internal double integrals are equal
$$
\int\limits_{-\infty}^{\infty}\int\limits_{-\infty}^{\infty}
\dfrac{dP_ydP_z}{1+\exp\Big(\Big(P_x\mp\dfrac{\hbar k}{p_T}\Big)^2+
P_y^2+P_z^2-\alpha\Big)}=
$$
$$
=\pi\ln\big(1+e^{-(P_x\mp\frac{\hbar k}{p_t})^2+\alpha}\big).
$$

Hence, the second integral is equal
$$
\int\dfrac{f_F^{++}-f_F^{--}}{\omega-v_T{kP_x}}P_xd^3P=
$$
$$
=\pi\int\limits_{-\infty}^{\infty}
\dfrac{\ln(1+e^{-(\tau-\frac{\hbar k}{p_T})^2
+\alpha})-\ln(1+e^{-(\tau+\frac{\hbar k}{p_T})^2+\alpha})}
{\omega-v_Tk\tau}d\tau=
$$
$$
=\pi\int\limits_{-\infty}^{\infty}
\ln\dfrac{1+\exp\Big[-\Big(\tau-\dfrac{\hbar k}{p_T}\Big)^2+\alpha\Big]}
{1+\exp\Big[-\Big(\tau+\dfrac{\hbar k}{p_T}\Big)^2+\alpha\Big]}
\dfrac{d\tau}{\omega-v_Tk\tau}.
$$

Thus, the quantity of the generated longitudinal current in
quantum plasma is equal
$$
j_{x}^{\rm long}(x,t)=
\dfrac{\pi e^3p_T^4A^2(x,t)}{2(2\pi\hbar)^3mc^2\hbar}\times
$$
$$
\times\Bigg[2\dfrac{mv_T^2}{\hbar}
\int\limits_{-\infty}^{\infty}\dfrac{l_1^{++}(\tau,\alpha)+
l_1^{--}(\tau,\alpha)-2l_1(\tau,\alpha)}{(\omega-v_Tk\tau )^2}\tau d\tau+
$$
$$
+\int\limits_{-\infty}^{\infty}
\ln\dfrac{1+\exp\Big[-\Big(\tau-\dfrac{\hbar k}{p_T}\Big)^2+\alpha\Big]}
{1+\exp\Big[-\Big(\tau+\dfrac{\hbar k}{p_T}\Big)^2+\alpha\Big]}
\dfrac{d\tau}{\omega-v_Tk\tau}\Bigg].
\eqno{(5.3)}
$$

At calculation  integrals from (5.3) it is necessary to take advantage
known Landau' rule, having included in an integration contour
semicircle with radius $ \varepsilon $ (laying in the bottom semiplane)
concerning of the pole $x_0 =\omega/(v_Tk) $, with the subsequent
transition to the limit at $ \varepsilon\to 0$. The first integral from
(5.3) it is necessary to integrate preliminary in parts.

Landau' rule completely is equivalent to following reception.
The pole is shifted from the real axis in the top semiplane in the point
$x_0 =\omega/(v_Tk)+i\varepsilon $. Then we integrate in parts and we pass to
limit at $ \varepsilon\to 0$. Not to repeat this reception each time,
let us carry out this calculation in a general view

$$
\int\limits_{a}^{b}\dfrac{\varphi(\tau)d\tau}{(\tau-x)^2}=
\lim\limits_{\varepsilon\to 0}\int\limits_{a}^{b}\dfrac{\varphi(\tau)d\tau}
{[\tau-(x+i\varepsilon)]^2}=
$$
$$
=\lim\limits_{\varepsilon\to 0}
\Big[-\dfrac{\varphi(\tau)}{\tau-(x+i\varepsilon)}\Bigg|_{a}^{b}+
\int\limits_{a}^{b}\dfrac{\varphi'(\tau)d\tau}{\tau-(x+i\varepsilon)}\Big]=
$$
$$
=-\dfrac{\varphi(\tau)}{\tau-x}\Bigg|_a^b+i\pi \varphi'(x)+
\int\limits_{a}^{b}\dfrac{\varphi'(\tau)d\tau}{\tau-x}.
$$

Last integral is understood in sense of a principal value (symbol V.P. we do
not write).

To the same result it is possible to come and as follows
$$
\int\limits_{a}^{b}\dfrac{\varphi(\tau)d\tau}{(\tau-x)^2}=
\lim\limits_{\varepsilon\to 0}\int\limits_{a}^{b}\dfrac{\varphi(\tau)d\tau}
{[\tau-(x+i\varepsilon)]^2}=
$$

$$
=\dfrac{d}{dx}\Bigg[
\lim\limits_{\varepsilon\to 0}\int\limits_{a}^{b}\dfrac{\varphi(\tau)d\tau}
{\tau-(x+i\varepsilon)}\Bigg]=\dfrac{d}{dx}\Big[i\pi \varphi(x)+
\int\limits_{a}^{b}\dfrac{\varphi(\tau)d\tau}{\tau-x}\Big]=
$$
$$
=i\pi\varphi'(x)-\dfrac{\varphi(\tau)}{\tau-x}\Bigg|_a^b+
\int\limits_{a}^{b}\dfrac{\varphi'(\tau)}{\tau-x}d\tau.
$$

We will designate
$$
f(\tau,\alpha)=l_1^{++}(\tau,\alpha)+l_1^{--}(\tau,\alpha)-2l_1(\tau,\alpha)
$$
and
$$
\varphi(\tau,\alpha)=\ln\dfrac{1+\exp\Big[-\Big(\tau-\dfrac{\hbar k}{p_T}\Big)^2
+\alpha\Big]}{1+\exp\Big[-\Big(\tau+\frac{\hbar k}{p_T}\Big)^2+\alpha\Big]}.
$$

Let us transform equality (5.3) in the form
$$
j_{x}^{\rm long}(x,t)=
\dfrac{\pi e^3p_T^4A^2(x,t)}{2(2\pi\hbar)^3mc^2\hbar}
\Bigg[\dfrac{2m}{\hbar k^2}\int\limits_{-\infty}^{\infty}
\dfrac{\tau f(\tau,\alpha)d\tau}{(\tau-x_0)^2}-\dfrac{1}{v_Tk}
\int\limits_{-\infty}^{\infty}\dfrac{\varphi(\tau,\alpha)d\tau}
{\tau-x_0}\Bigg],
$$
where
$$
x_0=\dfrac{\omega}{v_Tk}.
$$

Thus, according to the previous remark the
generated density of longitudinal current is equal
$$
j_{x}^{\rm long}(x,t)=
\dfrac{\pi e^3p_T^4A^2(x,t)}{2(2\pi\hbar)^3mc^2\hbar}
\Bigg[\dfrac{2m}{\hbar k^2}\int\limits_{-\infty}^{\infty}
\dfrac{(\tau f(\tau,\alpha))'d\tau}{\tau-x_0}+
$$
$$
+i\pi
\dfrac{\partial (\tau f(\tau,\alpha))}{\partial \tau}\Bigg|_{\tau=x_0}
-\dfrac{1}{v_Tk}
\int\limits_{-\infty}^{\infty}\dfrac{\varphi(\tau,\alpha)d\tau}
{\tau-x_0}\Bigg].
\eqno{(5.4)}
$$

\begin{center}
  \bf 6. Degenerate plasmas
\end{center}

Let's consider the case of degenerate plasmas.

In the formula (5.2) we will carry out replacement of one
variable of integration
$$
P_x\to \dfrac{v_F}{v_T}P_x,
$$
where $v_F$ is the electron velocity on Fermi' surface.

Let us receive following expression for quantity of electric
current density into quantum plasmas

$$
j_{x}^{\rm long}(x,t)=
\dfrac{e^3p_F^4A^2(x,t)}{2(2\pi\hbar)^3m^2c^2\hbar}\times
$$
$$
\times \Bigg[2\dfrac{mv_F^2}{\hbar}\int
\dfrac{f_F^{++}+f_F^{--}-2f_F}{(\omega-v_FkP_x)^2}P_xP_y^2d^3P+
\int\dfrac{f_F^{++}-f_F^{--}}{\omega-v_F{kP_x}}P_xd^3P\Bigg].
\eqno{(6.1)}
$$

In the formula (6.1) $p_F=mv_F$ is the electron momentum on Fermi'
surface,
$$
f_F=\dfrac{1}{1+\exp\dfrac{\E_FP^2-\mu}{k_BT}},
$$
where $\E_F$ is the electron energy on Fermi' surface,
$$
f_F^{\pm\pm}=\dfrac{1}{1+\exp\dfrac{\E_F\Big[\Big(P_x\mp
\dfrac{\hbar k}{p_F}\Big)^2+P_y^2+P_z^2\Big]-\mu}{k_BT}}.
$$

Let us notice, that in the limit of zero absolute temperature we have

$$
 \lim\limits_{T\to 0}\mu =\E_F, \qquad \E_F=\dfrac{mv_F^2}{2},
$$

Hence, in limit of zero temperature absolute Fermi---Dirac'
distribution passes in Fermi's absolute distribution for degenerate
plasmas
$$
\lim\limits_{T\to 0}f_F=\lim\limits_{T\to 0}\dfrac{1}{1+\exp\Big(
\dfrac{\E_FP^2-\mu}{k_BT}\Big)}=
$$
$$
=\Theta(\E_F(1-P^2))=
\Theta(1-P^2)=\Theta.
$$

Here $\Theta(x)$ is the Heaviside function (unit step),
$$
\Theta(x)=\Bigg\{ \begin{array}{l}
                   1, \quad x>0, \\
                   0,\quad x<0
                 \end{array}\Bigg\}.
$$

Similarly,
$$
\lim\limits_{T\to 0}f_F^{++}=\Theta\Big[1-\Big(P_x-
\dfrac{\hbar k}{p_F}\Big)^2-P_y^2-P_z^2\Big]=\Theta^{++},
$$
$$
\lim\limits_{T\to 0}f_F^{--}=\Theta\Big[1-\Big(P_x+
\dfrac{\hbar k}{p_F}\Big)^2-P_y^2-P_z^2\Big]=\Theta^{--}.
$$

Thus, in the limit of zero temperature the formula (6.1)
will be transformed to the form
$$
j_{x}^{\rm long}(x,t)=
\dfrac{e^3p_F^4A^2(x,t)}{2(2\pi\hbar)^3m^2c^2\hbar}\times
$$
$$
\times \Bigg[2\dfrac{mv_F^2}{\hbar}\int
\dfrac{\Theta^{++}+\Theta^{--}-2\Theta}{(\omega-v_FkP_x)^2}P_xP_y^2d^3P+
\int\dfrac{\Theta^{++}-\Theta^{--}}{\omega-v_F{kP_x}}P_xd^3P\Bigg].
\eqno{(6.2)}
$$

We note that
$$
\Big(\omega-kv_FP_x\Big)^2=(kv_F)^2\Big(P_x-\dfrac{\omega}{kv_F}\Big)^2,
$$
$$
\omega-kv_FP_x=-kv_F\Big(P_x-\dfrac{\omega}{kv_F}\Big).
$$

By means of these equalities we will transform the formula (6.2) to the
following  form
$$
j_{x}^{\rm long}(x,t)=
\dfrac{e^3p_F^2A^2(x,t)}{(2\pi\hbar)^3mc^2q^2}\times
$$
$$
\times \Bigg[\int
\dfrac{\Theta^{++}+\Theta^{--}-2\Theta}{(P_x-\omega/kv_F)^2}P_xP_y^2d^3P-
\dfrac{q}{2}
\int\dfrac{\Theta^{++}-\Theta^{--}}{P_x-\omega/kv_F}P_xd^3P\Bigg].
\eqno{(6.3)}
$$

In (6.3) Fermi's wave number $k_F $ and dimensionless wave number
$q$ is entered.

We consider the first integral from  (6.3). We have
$$
\int
\dfrac{\Theta^{++}+\Theta^{--}-2\Theta}{(P_x-\omega/kv_F)^2}P_xP_y^2d^3P=
$$
$$
=\int\dfrac{\Theta\Big[1-\Big(P_x-
\dfrac{\hbar k}{p_F}\Big)^2-P_y^2-P_z^2\Big]}
{(P_x-\omega/kv_F)^2}P_xP_y^2d^3P+
$$
$$
+\int\dfrac{\Theta\Big[1-\Big(P_x+
\dfrac{\hbar k}{p_F}\Big)^2-P_y^2-P_z^2\Big]}
{(P_x-\omega/kv_F)^2}P_xP_y^2d^3P-
$$
$$
-2\int\dfrac{\Theta(1-P^2)}
{(P_x-\omega/kv_F)^2}P_xP_y^2d^3P=J_++J_--2J_0.
$$

Here the following designations are entered

$$
J_+=\int\dfrac{\Theta\Big[1-\Big(P_x-
\dfrac{\hbar k}{p_F}\Big)^2-P_y^2-P_z^2\Big]}
{(P_x-\omega/kv_F)^2}P_xP_y^2d^3P=
$$
$$
=\int\limits_{\big(P_x-\frac{k\hbar}{p_F}\big)^2+P_y^2+P_z^2<1}
\dfrac{P_xP_y^2d^3P}{(P_x-\omega/kv_F)^2},
$$
$$
J_-=\int\dfrac{\Theta\Big[1-\Big(P_x+
\dfrac{\hbar k}{p_F}\Big)^2-P_y^2-P_z^2\Big]}
{(P_x-\omega/kv_F)^2}P_xP_y^2d^3P=
$$
$$
=\int\limits_{\big(P_x+\frac{k\hbar}{p_F}\big)^2+P_y^2+P_z^2<1}
\dfrac{P_xP_y^2d^3P}{(P_x-\omega/kv_F)^2},
$$
$$
J_0=\int\dfrac{\Theta(1-P^2)}
{(P_x-\omega/kv_F)^2}P_xP_y^2d^3P=
$$
$$
=\int\limits_{P_x^2+P_y^2+P_z^2<1}
\dfrac{P_xP_y^2d^3P}{(P_x-\omega/kv_F)^2}.
$$

In the integral $J_+$ we carry out replacement of  variable of
integration
$P_x-\dfrac{\hbar k}{p_F}\to P_x$, and in integral $J_-$ we replace
$P_x+\dfrac{\hbar k}{p_F}\to P_x$. As result we receive, that

$$
J_{+}=\int\limits_{P^2<1}\dfrac{\Big(P_x+\dfrac{\hbar k}{p_F}\Big)
P_y^2d^3P}{\Big[P_x+\dfrac{\hbar k}{p_F}-\dfrac{\omega}{kv_F}\Big]^2},
$$

$$
J_{-}=\int\limits_{P^2<1}\dfrac{\Big(P_x-\dfrac{\hbar k}{p_F}\Big)
P_y^2d^3P}{\Big[P_x-\dfrac{\hbar k}{p_F}-\dfrac{\omega}{kv_F}\Big]^2},
$$

$$
J_0=\int\limits_{P^2<1}\dfrac{P_xP_y^2d^3P}
{\Big[P_x-\dfrac{\omega}{kv_F}\Big]^2}.
$$

Let us reduce these integrals to the repeated

$$
J_{+}=\int\limits_{-1}^{1}\dfrac{\Big(P_x+\dfrac{\hbar k}{p_F}\Big)
dP_x}{\Big[P_x+\dfrac{\hbar k}{p_F}-\dfrac{\omega}{kv_F}\Big]^2}
\int\limits_{P_y^2+P_z^2<1-P_x^2}P_y^2dP_ydP_z,
$$

$$
J_{-}=\int\limits_{-1}^{1}\dfrac{\Big(P_x-\dfrac{\hbar k}{p_F}\Big)
dP_x}{\Big[P_x-\dfrac{\hbar k}{p_F}-\dfrac{\omega}{kv_F}\Big]^2}
\int\limits_{P_y^2+P_z^2<1-P_x^2}P_y^2dP_ydP_z,
$$

$$
J_0=\int\limits_{-1}^{1}\dfrac{P_x dP_x}
{\Big[P_x-\dfrac{\omega}{kv_F}\Big]^2}
\int\limits_{P_y^2+P_z^2<1-P_x^2}P_y^2dP_ydP_z.
$$

We will calculate internal double integrals, passing to the polar
coor\-di\-nates
$$
P_y=\rho\cos \varphi,\quad P_z=\rho\sin\varphi, \quad 0\leqslant \varphi
\leqslant 2\pi, \quad 0\leqslant \rho \leqslant \sqrt{1-P_x^2}.
$$

We receive, that

$$
\int\limits_{P_y^2+P_z^2<1-P_x^2}P_y^2dP_ydP_z=\int\limits_{0}^{2\pi}
\cos^2\varphi d\varphi
\int\limits_{0}^{\sqrt{1-P_x^2}}\rho^3 d\rho=\dfrac{\pi}{4}(1-P_x^2)^2.
$$

Thereforte, we receive the following one-dimensional integrals
$$
J_{+}=\dfrac{\pi}{4}\int\limits_{-1}^{1}
\dfrac{\Big(P_x+\dfrac{\hbar k}{p_F}\Big)(1-P_x^2)^2dP_x}
{\Big[P_x+\dfrac{\hbar k}{p_F}-\dfrac{\omega}{kv_F}\Big]^2},
$$
$$
J_{-}=\dfrac{\pi}{4}\int\limits_{-1}^{1}
\dfrac{\Big(P_x-\dfrac{\hbar k}{p_F}\Big)(1-P_x^2)^2dP_x}
{\Big[P_x-\dfrac{\hbar k}{p_F}-\dfrac{\omega}{kv_F}\Big]^2},
$$
$$
J_0=\dfrac{\pi}{4}\int\limits_{-1}^{1}\dfrac{P_x(1-P_x^2)^2dP_x}
{\Big[P_x-\dfrac{\omega}{kv_F}\Big]^2}.
$$

So, the first integral from(6.3) equals
$$
\int
\dfrac{\Theta^{++}+\Theta^{--}-2\Theta}{(P_x-\omega/kv_F)^2}P_xP_y^2d^3P=
\dfrac{\pi}{4}\int\limits_{-1}^{1}
\dfrac{\Big(P_x+\dfrac{\hbar k}{p_F}\Big)(1-P_x^2)^2dP_x}
{\Big[P_x+\dfrac{\hbar k}{p_F}-\dfrac{\omega}{kv_F}\Big]^2}+
$$
$$
+\dfrac{\pi}{4}\int\limits_{-1}^{1}
\dfrac{\Big(P_x-\dfrac{\hbar k}{p_F}\Big)(1-P_x^2)^2dP_x}
{\Big[P_x-\dfrac{\hbar k}{p_F}-\dfrac{\omega}{kv_F}\Big]^2}-2
\dfrac{\pi}{4}\int\limits_{-1}^{1}\dfrac{P_x(1-P_x^2)^2dP_x}
{\Big[P_x-\dfrac{\omega}{kv_F}\Big]^2}.
$$

The second integral from (6.3) is calculated similarly

$$
\int\dfrac{\Theta^{++}-\Theta^{--}}{P_x-{\omega}/{kv_F}}P_xd^3P=
$$
$$
=\pi\int\limits_{-1}^{1}\dfrac{(1-P_x^2)\Big(P_x+\dfrac{\hbar k}{p_F}\Big)dP_x}
{P_x+\dfrac{\hbar k}{p_F}-\dfrac{\omega}{kv_F}}-
\pi\int\limits_{-1}^{1}\dfrac{(1-P_x^2)\Big(P_x-\dfrac{\hbar k}{p_F}\Big)dP_x}
{P_x-\dfrac{\hbar k}{p_F}-\dfrac{\omega}{kv_F}}.
$$

Hence, the generated longitudinal electric current is equal

$$
j_{x}^{\rm long}(x,t)=
\dfrac{e^3p_F^2A^2(x,t)}{32\pi^2\hbar^3mc^2q^2}\Bigg\{
\Bigg[\int\limits_{-1}^{1}
\dfrac{(1-\tau^2)^2(\tau+\tau_0)d\tau}{(\tau+\tau_0-x_0)^2}+
$$
$$
+\int\limits_{-1}^{1}
\dfrac{(1-\tau^2)^2(\tau-\tau_0)d\tau}{(\tau-\tau_0-x_0)^2}
-2\int\limits_{-1}^{1}
\dfrac{(1-\tau^2)^2\tau d\tau}{(\tau-x_0)^2}\Bigg]-
$$
$$
-2q\Bigg[\int\limits_{-1}^{1}\dfrac{(1-\tau^2)
(\tau+\tau_0)d\tau}{\tau+\tau_0-x_0}-\int\limits_{-1}^{1}
\dfrac{(1-\tau^2)(\tau-\tau_0)d\tau}{\tau-\tau_0-x_0}\Bigg]\Bigg\}.
\eqno{(6.4)}
$$

In (6.4) the  following designations are accepted
$$
\tau=P_x,\qquad \tau_0=\dfrac{\hbar k}{p_F},\qquad x_0=\dfrac{\omega}{v_Fk}.
$$

First two integrals from (6.4) we will present as one equality
$$
J^{\pm}=\int\limits_{-1}^{1}\dfrac{(1-\tau^2)^2(\tau\pm\tau_0)d\tau}
{(\tau\pm\tau_0-x_0)^2}=\int\limits_{-1}^{1}
\dfrac{[(1-\tau^2)^2(\tau\pm\tau_0)]'d\tau}{\tau\pm\tau_0-x_0}+
$$
$$
+i\pi [(1-\tau^2)^2(\tau\pm\tau_0)]'\Bigg|_{\tau=x_0\mp\tau_0}\times
\Big\{\begin{array}{l}
         1,\quad |x_0\mp\tau_0|<1 \\
         0,\quad |x_0\mp\tau_0|>1       
\end{array}\Big\}.
\eqno{(6.5)}
$$

In the same way the third integral is equal
$$
J^0=\int\limits_{-1}^{1}\dfrac{\tau(1-\tau^2)^2d\tau}{(\tau-x_0)^2}=
\int\limits_{-1}^{1}\dfrac{[\tau(1-\tau^2)^2]'d\tau}{\tau-x_0}+
$$
$$
+i\pi [\tau(1-\tau^2)^2]'\Bigg|_{\tau=x_0}\times
\Big\{\begin{array}{l}
         1,\quad |x_0|<1 \\
         0,\quad |x_0|>1
       \end{array}\Big\}.
\eqno{(6.6)}
$$

Let us calculate in an explicit form integral from (6.6)
$$
J^0=\int\limits_{-1}^{1}\dfrac{[\tau(1-\tau^2)^2]'d\tau}{\tau-x_0}=
10x_0^2-\dfrac{26}{3}x_0+
$$
$$
+(5x_0^4-6x_0^2+1)\Big[\ln\Big|\dfrac{x_0-1}{x_0+1}\Big|+
\Big\{\begin{array}{r}
         i\pi,\quad |x_0|<1 \\
         0,\quad |x_0|>1
       \end{array}\Big\}\Big].
\eqno{(6.7)}
$$

Let us pass to calculation of integrals from equality (6.5). We have
$$
\int\limits_{-1}^{1}
\dfrac{[(1-\tau^2)^2(\tau+\tau_0)]'d\tau}{\tau+\tau_0-x_0}=
$$
$$
=\int\limits_{-1}^{1}
\dfrac{1-4\tau_0\tau-6\tau^2+4\tau_0\tau^3+5\tau^4}{\tau+\tau_0-x_0}d\tau.
$$

After replacement of a variable of integration
$$
u=\tau+\tau_0-x_0,\quad u_0=-1+\tau_0-x_0,\quad u_1=1+\tau_0-x_0,
$$
we receive
$$
\int\limits_{u_0}^{u_1}\Big[1-4\tau_0(u-\tau_0+x_0)-6(u-\tau_0+
x_0)^2+
$$
$$
+4\tau_0(u-\tau_0+x_0)^3+5(u-\tau_0+x_0)^4\Big]\dfrac{du}{u}=
$$
$$
=\int\limits_{u_0}^{u_1}\Bigg\{\Big[1-4\tau_0(x_0-\tau_0)-
6(x_0-\tau_0)^2+4\tau_0(x_0-\tau_0)^3+5(x_0-\tau_0)^4\Big]+
$$
$$
+\Big[-4\tau_0-12(x_0-\tau_0)+12\tau_0(x_0-\tau_0)^2+
20(x_0-\tau_0)^3\Big]u+
$$
$$
+\Big[-6+12\tau_0(x_0-\tau_0)+30(x_0-\tau_0)^2\Big]u^2+
\Big[4\tau_0+20(x_0-\tau_0)\Big]u^3+5u^4\Bigg\}\dfrac{du}{u}.
$$

As a result of rectilinear calculations we receive, that
$$
J^+=
\Big[5(x_0-\tau_0)^4+4\tau_0(x_0-\tau_0)^3-6(x_0-\tau_0)^2-4\tau_0(x_0-\tau_0)
+1\Big]\times
$$\medskip
$$
\times\Big[\ln\dfrac{x_0-\tau_0-1}{x_0-\tau_0+1}+
\Big\{\begin{array}{r}
         i\pi,\quad |x_0-\tau_0|<1 \\
         0,\quad |x_0-\tau_0|>1
       \end{array}\Big\}\Big]+\dfrac{2}{3}[5(x_0-\tau_0)+4\tau_0]+
$$\medskip
$$
+2[5(x_0-\tau_0)^3+4\tau_0(x_0-\tau_0)^2-6(x_0-\tau_0)-4\tau_0].
\eqno{(6.8)}
$$\medskip

Let us calculate the second integral from (6.5). We have
$$
J^-=
\Big[5(x_0+\tau_0)^4-4\tau_0(x_0+\tau_0)^3-6(x_0+\tau_0)^2+4\tau_0(x_0+\tau_0)
+1\Big]\times
$$
$$
\times\Big[\ln\Big|\dfrac{x_0+\tau_0-1}{x_0+\tau_0+1}\Big|+
\Big\{\begin{array}{r}
         i\pi,\quad |x_0+\tau_0|<1 \\
         0,\quad |x_0+\tau_0|>1
       \end{array}\Big\}\Big]+\dfrac{2}{3}[5(x_0+\tau_0)-4\tau_0]+
$$
$$
+2[5(x_0+\tau_0)^3-4\tau_0(x_0+\tau_0)^2-6(x_0+\tau_0)+4\tau_0].
\eqno{(6.9)}
$$ \medskip

We notice that from equalities (6.8) and (6.9) follows that
$$
J^+\Big|_{\tau_0=0}=J^-\Big|_{\tau_0=0}=J^0.
$$

Besides, we will notice, that
$$
5x_0^4-6x_0^2+1=(x_0^2-1)(5x_0^2-1),
$$\medskip
$$
5(x_0-\tau_0)^4+4\tau_0(x_0-\tau_0)^3-6(x_0-\tau_0)^2-4\tau_0(x_0-\tau_0)
+1=
$$
$$
=[(x_0-\tau_0)^2-1][5(x_0-\tau_0)^2+4\tau_0(x_0-\tau_0)-1],
$$ \medskip
$$
5(x_0+\tau_0)^4-4\tau_0(x_0+\tau_0)^3-6(x_0+\tau_0)^2+4\tau_0(x_0+\tau_0)
+1=
$$
$$
=[(x_0+\tau_0)^2-1][5(x_0+\tau_0)^2-4\tau_0(x_0+\tau_0)-1].
$$ \medskip

The first square bracket from (6.4) thanking (6.7) -- (6.9) is equal
$$
J^++J^--2J^0=28x_0\tau_0^2+
$$
$$
+[(x_0-\tau_0)^2-1][5(x_0-\tau_0)^2+4\tau_0(x_0-\tau_0)-1] \times
$$
$$
\times\Big[\ln\Big|\dfrac{x_0-\tau_0-1}{x_0-\tau_0+1}\Big|+
\Big\{\begin{array}{r}
         i\pi,\quad |x_0-\tau_0|<1 \\
         0,\quad |x_0-\tau_0|>1
       \end{array}\Big\}\Big]+
$$
$$
+[(x_0+\tau_0)^2-1][5(x_0+\tau_0)^2-4\tau_0(x_0+\tau_0)-1] \times
$$
$$
\times\Big[\ln\Big|\dfrac{x_0+\tau_0-1}{x_0+\tau_0+1}\Big|+
\Big\{\begin{array}{r}
         i\pi,\quad |x_0+\tau_0|<1 \\
         0,\quad |x_0+\tau_0|>1
       \end{array}\Big\}\Big]-
$$
$$
-2(x_0^2-1)(5x_0^2-1)\Big[\ln\Big|\dfrac{x_0-1}{x_0+1}\Big|+
\Big\{\begin{array}{r}
         i\pi,\quad |x_0|<1 \\
         0,\quad |x_0|>1
       \end{array}\Big\}\Big].
$$

The remained two integrals from equality (6.3) are calculated or
by means of Landau' rule or the same as it has been stated above.
The first of these integrals is calculated as follows
$$
I^+ \equiv \int\limits_{-1}^{1}\dfrac{(1-\tau^2)(\tau+\tau_0)}{\tau-
(x_0-\tau_0)}d\tau=\lim\limits_{\varepsilon\to 0}
\int\limits_{-1}^{1}\dfrac{(1-\tau^2)(\tau+\tau_0)}{\tau-
(x_0+i\varepsilon-\tau_0)}d\tau+
$$
$$
+i\pi[1-(x_0-\tau_0)^2]x_0\Big\{\begin{array}{l}
                                  1,|x_0-\tau_0|<1 \\
                                  0, |x_0-\tau_0|>1
                                \end{array}
\Big\}+
$$
$$
+ \int\limits_{-1}^{1}\dfrac{(1-\tau^2)(\tau+\tau_0)}{\tau-
(x_0-\tau_0)}d\tau,
$$
and last integral is understood in sense of a principal value.
This integral is equal
$$
\int\limits_{-1}^{1}\dfrac{(1-\tau^2)(\tau+\tau_0)}{\tau-
(x_0-\tau_0)}d\tau=\dfrac{4}{3}-2\tau_0(x_0-\tau_0)-2(x_0-\tau_0)^2+
$$
$$
+\Big[x_0-\tau_0(x_0-\tau_0)^2-(x_0-\tau_0)^3\Big]\ln\Big|\dfrac{x_0-\tau_0-1}
{x_0-\tau_0+1}\Big|.
$$

The second integral is calculated similarly. We have
$$
I^- \equiv \int\limits_{-1}^{1}\dfrac{(1-\tau^2)(\tau-\tau_0)}{\tau-
(x_0+\tau_0)}d\tau=\lim\limits_{\varepsilon\to 0}
\int\limits_{-1}^{1}\dfrac{(1-\tau^2)(\tau-\tau_0)}{\tau-
(x_0+i\varepsilon+\tau_0)}d\tau=
$$
$$
=i\pi[1-(x_0+\tau_0)^2]x_0\Big\{\begin{array}{l}
                                  1,|x_0+\tau_0|<1 \\
                                  0, |x_0+\tau_0|>1
                                \end{array}
\Big\}+
$$
$$
+ \int\limits_{-1}^{1}\dfrac{(1-\tau^2)(\tau-\tau_0)}{\tau-
(x_0+\tau_0)}d\tau,
$$
and last integral is understood in sense of a principal value.
This integral is equal
$$
\int\limits_{-1}^{1}\dfrac{(1-\tau^2)(\tau-\tau_0)}{\tau-
(x_0+\tau_0)}d\tau=\dfrac{4}{3}+2\tau_0(x_0-\tau_0)-2(x_0-\tau_0)^2+
$$
$$
+\Big[x_0+\tau_0(x_0+\tau_0)^2-(x_0+\tau_0)^3\Big]\ln\Big|\dfrac{x_0+\tau_0-1}
{x_0+\tau_0+1}\Big|.
$$

We notice that
$$
-(x_0-\tau_0)^3-\tau_0(x_0-\tau_0)+x_0=x_0[1-(x_0-\tau_0)^2],
$$
$$
-(x_0+\tau_0)^3+\tau_0(x_0+\tau_0)+x_0=x_0[1-(x_0+\tau_0)^2].
$$

Now we will find the difference $I^+-I^-$. We have
$$
I^+-I^-=4x_0\tau_0+
$$
$$
+x_0[1-(x_0-\tau_0)^2]\Big[\ln\Big|\dfrac{x_0-\tau_0-1}{x_0-\tau_0+1}\Big|+
\Big\{\begin{array}{r}
                                  i\pi,|x_0-\tau_0|<1 \\
                                  0, |x_0-\tau_0|>1
                                \end{array}\Big\}\Big]-
$$
$$
-x_0[1-(x_0+\tau_0)^2]
\Big[\ln\Big|\dfrac{x_0+\tau_0-1}{x_0+\tau_0+1}\Big|+
\Big\{\begin{array}{r}
                                  i\pi,|x_0+\tau_0|<1 \\
                                  0, |x_0+\tau_0|>1
                                \end{array}\Big\}\Big].
$$

It is necessary to find density electric current  under the formula
$$
j_{x}^{\rm long}(x,t)=
\dfrac{e^3p_F^2A^2(x,t)}{32\pi^2\hbar^3mc^2q^2}\Big[(J^++J^--2J^0)
-2q(I^+-I^-)\Big].
$$

Substituting the found integrals in this equality, we receive

$$
j_{x}^{\rm long}(x,t)=
\dfrac{e^3p_F^2A^2(x,t)}{32\pi^2\hbar^3mc^2q^2}\Bigg[20x_0\tau_0^2+
$$
$$
+[(x_0-\tau_0)^2-1]\Big[5(x_0-\tau_0)^2+4\tau_0(x_0-\tau_0)+
2x_0\tau_0-1\Big] \times
$$
$$
\times\Big[\ln\Big|\dfrac{x_0-\tau_0-1}{x_0-\tau_0+1}\Big|+
\Big\{\begin{array}{r}
         i\pi,\quad |x_0-\tau_0|<1 \\
         0,\quad |x_0-\tau_0|>1
       \end{array}\Big\}\Big]+
$$
$$
+[(x_0+\tau_0)^2-1]\Big[5(x_0+\tau_0)^2-4\tau_0(x_0+\tau_0)-
2x_0\tau_0-1\Big] \times
$$
$$
\times\Big[\ln\Big|\dfrac{x_0+\tau_0-1}{x_0+\tau_0+1}\Big|+
\Big\{\begin{array}{r}
         i\pi,\quad |x_0+\tau_0|<1 \\
         0,\quad |x_0+\tau_0|>1
       \end{array}\Big\}\Big]-
$$
$$
-2(x_0^2-1)(5x_0^2-1)\Big[\ln\Big|\dfrac{x_0-1}{x_0+1}\Big|+
\Big\{\begin{array}{r}
         i\pi,\quad |x_0|<1 \\
         0,\quad |x_0|>1
       \end{array}\Big\}\Big].
\Bigg]
\eqno{(6.10)}
$$

We will remind that $q=\dfrac{k}{k_F}$ is the dimensionless wave
number, where $k_F$ is the wave Fermi' number,
$k_F=\dfrac{p_F}{\hbar}$, $p_F=mv_F$ is the electron momentum
on Fermi' surface,
$$
\tau_0=\dfrac{\hbar k}{mv_F}=\dfrac{k}{k_F}=q,\qquad
x_0=\dfrac{\omega}{kv_F}=\dfrac{\omega}{v_Fk_F}\cdot\dfrac{1}{q}=
\dfrac{\Omega}{q},
$$
where
$$
\Omega=\dfrac{\omega}{v_Fk_F}.
$$

Now  it is possible to present the formula (6.10) in the form

$$
j_{x}^{\rm long}(x,t)=
\dfrac{e^3p_F^2A^2(x,t)}{32\pi^2\hbar^3mc^2q^2}\Bigg[20\Omega q+
\Big[\Big(\dfrac{\Omega}{q}-q\Big)^2-1\Big]
\Big[5\dfrac{\Omega^2}{q^2}-4\Omega+q^2-1\Big] \times
$$
$$
\times\Big[\ln\Big|\dfrac{\Omega-q^2-q}{\Omega-q^2+q}\Big|+
\Big\{\begin{array}{r}
i\pi,\quad |\Omega-q^2|<|q| \\0,\quad |\Omega-q^2|>|q|
\end{array}\Big\}\Big]+\Big[\Big(\dfrac{\Omega}{q}+q\Big)^2-1\Big]\times
$$
$$
\Big[5\dfrac{\Omega^2}{q^2}+4\Omega+q^2-1\Big]
\Big[\ln\Big|\dfrac{\Omega+q^2-q}{\Omega+q^2+q}\Big|+
\Big\{\begin{array}{r}
         i\pi,\quad |\Omega+q^2|<|q| \\
         0,\quad |\Omega+q^2|>|q|
       \end{array}\Big\}\Big]-
$$
$$
-2\Big(\dfrac{\Omega^2}{q^2}-1\Big)\Big(5\dfrac{\Omega^2}{q^2}-1\Big)
\Big[\ln\Big|\dfrac{\Omega-q}{\Omega+q}\Big|+\Big\{\begin{array}{r}
         i\pi,\quad |\Omega|<|q| \\
         0,\quad |\Omega|>|q|
       \end{array}\Big\}\Big].
\Bigg]
\eqno{(6.11)}
$$

Let us allocate at equality (6.11) on the right the real and imaginary parts
$$
j_x^{\rm long}(x,t)=\dfrac{e^3 p_F^2A^2(x,t)}{32\pi^2\hbar^3mc^2}
\dfrac{R(\Omega,q)+i\pi S(\Omega,q)}{q^2}.
\eqno{(6.12)}
$$

In expression (6.12) the following designations are entered
$$
R(\Omega,q)=20\Omega q+\Big[\Big(\dfrac{\Omega}{q}-q\Big)^2-1\Big]
\Big[5\dfrac{\Omega^2}{q^2}-4\Omega+q^2-1\Big] \times
$$
$$
\times\ln\Big|\dfrac{\Omega-q^2-q}{\Omega-q^2+q}\Big|+
\Big[\Big(\dfrac{\Omega}{q}+q\Big)^2-1\Big]
\Big[5\dfrac{\Omega^2}{q^2}+4\Omega+q^2-1\Big] \times
$$
$$
\times\ln\Big|\dfrac{\Omega+q^2-q}{\Omega+q^2+q}\Big|-2
\Big(\dfrac{\Omega^2}{q^2}-1\Big)\Big(5\dfrac{\Omega^2}{q^2}-1\Big)
\ln\Big|\dfrac{\Omega-q}{\Omega+q}\Big|,
$$
and

$$
S(\Omega,q)=\Big[\Big(\dfrac{\Omega}{q}-q\Big)^2-1\Big]
\Big[5\dfrac{\Omega^2}{q^2}-4\Omega+q^2-1\Big] \Big\{
\begin{array}{l}1,|\Omega-q^2|<|q| \\0,|\Omega-q^2|>|q|\end{array}\Big\} +
$$
$$
+\Big[\Big(\dfrac{\Omega}{q}+q\Big)^2-1\Big]
\Big[5\dfrac{\Omega^2}{q^2}+4\Omega+q^2-1\Big] \Big\{\begin{array}{l}
  1,|\Omega+q^2|<|q| \\0,|\Omega+q^2|>|q|\end{array}\Big\}-
$$
$$
-2\Big(\dfrac{\Omega^2}{q^2}-1\Big)\Big(5\dfrac{\Omega^2}{q^2}-1\Big)
\Big\{\begin{array}{l}1,|\Omega|<|q| \\
0,|\Omega|>|q|\end{array}\Big\}.
$$

\begin{center}
\bf 7. Conclusions
\end{center}

In the present work the account of nonlinear character of interaction
of electromagnetic field with quantum plasma is considered.
It has appeared, that
the effect of nonlinearity of an electromagnetic field affects in
generating of an electric current, transversal to a direction
of electromagnetic fields.

\end{document}